\documentclass[11pt,a4paper]{article} 
\usepackage{epsfig}
\usepackage{jheppub}
\usepackage[english]{babel}
\usepackage{graphicx}
\usepackage{color}
\usepackage{xspace}
\usepackage{placeins}
\usepackage{rotating}
\usepackage{bm}
\usepackage{booktabs}
\usepackage{amsmath}
\usepackage{indentfirst} 
\usepackage{longtable,multirow}
\usepackage{slashed}

\DeclareMathSymbol{\Gamma}{\mathalpha}{letters}{"00}
\DeclareMathSymbol{\Delta}{\mathalpha}{letters}{"01}
\DeclareMathSymbol{\Theta}{\mathalpha}{letters}{"02}
\DeclareMathSymbol{\Lambda}{\mathalpha}{letters}{"03}
\DeclareMathSymbol{\Xi}{\mathalpha}{letters}{"04}
\DeclareMathSymbol{\Pi}{\mathalpha}{letters}{"05}
\DeclareMathSymbol{\Sigma}{\mathalpha}{letters}{"06}
\DeclareMathSymbol{\Upsilon}{\mathalpha}{letters}{"07}
\DeclareMathSymbol{\Phi}{\mathalpha}{letters}{"08}
\DeclareMathSymbol{\Psi}{\mathalpha}{letters}{"09}
\DeclareMathSymbol{\Omega}{\mathalpha}{letters}{"0A}
\DeclareMathSymbol{\varGamma}{\mathalpha}{operators}{"00}
\DeclareMathSymbol{\varDelta}{\mathalpha}{operators}{"01}
\DeclareMathSymbol{\varTheta}{\mathalpha}{operators}{"02}
\DeclareMathSymbol{\varLambda}{\mathalpha}{operators}{"03}
\DeclareMathSymbol{\varXi}{\mathalpha}{operators}{"04}
\DeclareMathSymbol{\varPi}{\mathalpha}{operators}{"05}
\DeclareMathSymbol{\varSigma}{\mathalpha}{operators}{"06}
\DeclareMathSymbol{\varUpsilon}{\mathalpha}{operators}{"07}
\DeclareMathSymbol{\varPhi}{\mathalpha}{operators}{"08}
\DeclareMathSymbol{\varPsi}{\mathalpha}{operators}{"09}
\DeclareMathSymbol{\varOmega}{\mathalpha}{operators}{"0A}

\renewcommand{\vec}[1]{\boldsymbol{#1}}
\newcommand{\D}{\mathrm{d}}
\newcommand{\E}{\mathrm{e}}

\def\beq{\begin{equation}}
\def\eeq{\end{equation}}
\def\bea{\begin{eqnarray}}
\def\eea{\end{eqnarray}}
\def\bi{\begin{itemize}}
\def\ei{\end{itemize}}

\newcommand{\stau}{{\widetilde{\tau}}}

\newcommand{\neu}{{\widetilde{\chi}^0}}
\newcommand{\mstau}{m_{\stau_1}}

\newcommand{\mne}{{m_{\s{\chi}^0_1}}}
  
\newcommand{\charg}{{\s{\chi}^\pm_1}}

\newcommand{\sq}{{\widetilde{q}}}

\newcommand{\pt}{{p_\textnormal{T}}}

\newcommand{\GEV}{\ensuremath{\,\textnormal{GeV}}}
\newcommand{\TEV}{\ensuremath{\,\textnormal{TeV}}}

\newcommand{\s}[1]{\widetilde{#1}}

\newcommand{\smodels}{{\textsc{SModelS }}}

\newcommand{\omg}{\Omega_{\neu} h^2}

\title{Simplified Models for Exotic BSM Searches}

\author[1]{Jan Heisig,}
\author[2,3]{Andre Lessa,}
\author[4]{Loic Quertenmont}

\emailAdd{heisig@physik.rwth-aachen.de}
\emailAdd{andre.lessa@ufabc.edu.br}
\emailAdd{loic.quertenmont@gmail.com}

\affiliation[1]{
Institute for Theoretical Particle Physics and Cosmology,
\\ RWTH Aachen University, Germany
}

\affiliation[2]{
Instituto de F\'{i}sica, Universidade de S\~{a}o Paulo,
\\ R. do Mat\~{a}o 187, S\~{a}o Paulo, SP 05508-900, Brazil }

\affiliation[3]{Universidade Federal do ABC, \\
Santo Andr\'e, SP 09210-580, Brazil}

\affiliation[4]{Centre for Particle Physics and Phenomenology, CP3,\\
Universit\'e Catholique de Louvain, Belgium}

\abstract{ 
Simplified models are a successful way of interpreting current LHC searches for 
models beyond the standard model (BSM). So far simplified models have
focused on topologies featuring a missing transverse energy (MET) 
signature. However, in some BSM theories other, more exotic, signatures occur. 
If a charged particle becomes long-lived on collider time scales  -- as it is the 
case in parts of the SUSY parameter space --  it leads to a very
distinct signature.
We present an extension of the computer package SModelS which includes 
simplified models for heavy stable charged particles (HSCP). As a physical 
application we investigate the CMSSM stau co-annihilation strip 
containing long-lived staus, which presents a potential solution to the Lithium problem.
Applying both MET and HSCP constraints we show that, for low values of
$\tan\beta$, all this region of parameter space either violates Dark Matter
constraints or is excluded by LHC searches.}


\preprint{ 
TTK-15-20,
CP3-15-28
}

\arxivnumber{1509.00473}

\begin{document}

\maketitle

\section{Introduction}

In theories beyond the standard model (BSM) with an unbroken $\mathbb{Z}_2$-symmetry 
the lightest $\mathbb{Z}_2$-odd particle (LOP) is stable and hence usually required to be 
neutral as there exist strong bounds on the presence of stable charged particles in the 
universe~\cite{Smith:1982qu,Hemmick:1989ns,Yamagata:1993jq}. 
The typical collider signature of such a BSM scenario is missing transverse energy (MET) 
caused by the invisible LOP escaping the collider. However, there are scenarios where a 
heavy $\mathbb{Z}_2$-odd particle can become sufficiently long-lived to appear as
stable in a collider experiment. This particle can be charged and thus produce a very 
distinct signature.

Heavy stable charged particles (or HSCP) can appear when the next-to-lightest 
$\mathbb{Z}_2$-odd particle is nearly mass degenerate to the LOP and hence its decay
is kinematically suppressed. A prominent example for
such a situation is a supersymmetric scenario where a wino- or higgsino-like neutralino is the
lightest supersymmetric particle (LSP). The typically small mass splitting between the lightest 
neutralino and the lightest chargino in such a scenario can render the chargino long-lived, 
see e.g.~\cite{Bomark:2013nya}. A similar situation might occur in models of extra 
dimensions~\cite{Byrne:2003sa}. Other supersymmetric scenarios with mass
de\-ge\-ne\-rate sparticles have been proposed in order to address the discrepancy between 
the $^6$Li, $^7$Li abundances predicted in standard big bang nucleosynthesis (BBN) and 
those inferred from astrophysical observations \cite{Jittoh:2007fr}. In this scenario a 
bino-like neutralino and the lightest charged slepton are close in mass, resulting in a 
long-lived slepton~\cite{Jittoh:2005pq}.
Another possible scenario leading to HSCPs corresponds to the case
where the LOP does not share the SM gauge couplings
and only interacts super weakly. An important example are supersymmetric
scenarios where the gravitino or axino is the
LSP~\cite{Pagels:1981ke,Covi:1999ty}. In this case the couplings are suppressed
by powers of the Planck scale  or the Peccei-Quinn scale and the life-time of
the next-to-LSP (NLSP) can easily exceed the typical time for passing the
detectors by many orders of magnitude. 

Collider constraints on HSCPs have been mostly presented in specific BSM
mo\-dels~\cite{CMS1305.0491,Khachatryan:2015lla,ATLAS:2014fka} and in most
cases cannot be directly applied to other BSM scenarios. Currently, Simplified
Model Spectra (SMS)~\cite{Alwall:2008ag,Alves:2011wf,Okawa:2011xg,Chatrchyan:2013sza} 
have become a popular alternative for presenting less model-dependent constraints,
which can then be systematically applied to specific BSM scenarios.
The CMS and ATLAS collaborations typically interpret their results in
terms of simplified models and methods to use these interpretations
in a systematical way have been made publicly available for missing energy (MET)
topologies by tools such as \textsc{SModelS}~\cite{Kraml:2013mwa} and 
\textsc{Fastlim}~\cite{Papucci:2014rja}.
As a result it is now possible (under some approximations) to test general
BSM scenarios with MET signatures against LHC data.
So far a similar approach has not been considered for
HSCP or mixed MET-HSCP scenarios, although it has been
argued~\cite{Heisig:2012zq,Heisig:2013rya} that SMS are particularly suitable
for parametrizing the LHC sensitivity to HSCP signatures, since
these are rather inclusive and depend almost only on the kinematics of
the HSCP itself.

In this study we examine limits on the particle spectrum of BSM theories containing 
HSCPs making use of simplified models.
For this purpose we introduce a set of eight simplified model
topologies containing either one or two HSCPs in the final states and
 compute efficiencies for several values of the BSM masses appearing in each
topology. The resulting efficiency grids or {\it efficiency maps} allows us to
re-interpret previous results in a wide range of BSM models.
In order to do this in a more general framework we incorporate these efficiency
maps to the program package 
\textsc{SModelS}~\cite{Kraml:2013mwa,Kraml:2014sna}. Since the public version of
\textsc{SModelS} already contains a large number of LHC constraints on
simplified models containing MET signatures, our modified version allows us to
simultaneously apply both MET and HSCP constraints to full BSM models.
Hence we are able to test scenarios where HSCP and MET 
constraints compete.
As we will show, current LHC searches provide a high sensitivity to 
 HSCPs~\cite{Khachatryan:2015lla,ATLAS:2014fka} and
the HSCP signature can support an exclusion or discovery even if its
contribution to the total BSM signal is subdominant.

As a HSCP lead to a non-standard signature that is not supported by
common fast detector simulations particular attention has to be drawn to
a reliable implementation of such an analysis. In this study we use a novel 
approach for the computation of signal efficiencies presented in Ref.~\cite{Khachatryan:2015lla}. 
This method uses a parametrization of the CMS detector response as a function of 
the kinematic properties of the HSCP and allows us to accurately compute the CMS
signal efficiencies for arbitrary models.

As an application of the SMS framework to HSCPs, we consider
the stau co-annihilation strip in the Constrained Minimal 
Supersymmetric Standard Model (CMSSM). We focus on the
nearly mass degenerate neutralino LSP and stau NLSP,
which has been proposed in order to solve the $^7$Li
problem~\cite{Konishi:2013gda}.
Using our extension of \textsc{SModelS}, we study the
implications of both MET and HSCP searches and
show that all of the parameter space (with $\tan\beta=10$) consistent
with a potential solution to the $^7$Li problem and the observed
Dark Matter relic abundance is excluded by either HSCP or MET constraints (or both).
This region of the CMSSM parameter space has also been
studied in~\cite{Citron:2012fg,Desai:2014uha},
although not focussing on the region of interest for the
solution of the $^7$Li problem.
However, in~\cite{Desai:2014uha} no particular attention was drawn to the derivation
of the efficiencies for the HSCP signal and as an approximation the cross section limits from the 
inclusive stau production presented in~\cite{CMS1305.0491} were used. By applying 
the SMS framework to the considered slice of the CMSSM our derivation of the HSCP constraints
provide a significant improvement to the previous work.

The remainder of this paper is organized as follows. In Sec.~\ref{sec:defSMS}
we will define the simplified models used in our results and present
the computation of efficiency maps as well as their validation
against the results presented in the CMS
search~\cite{Khachatryan:2015lla}.
In Sec.~\ref{sec:sms} we explain our implementation 
of the decomposition of a full BSM model into simplified model topologies
and how these are used to constrain the full model.
An application of our results to a CMSSM scenario with a nearly 
mass degenerate neutralino and stau will be presented in 
Sec.~\ref{sec:appl}. We conclude in Sec.~\ref{sec:concl}.

\section{Simplified Models for HSCPs} \label{sec:defSMS}

In this section we will briefly review how CMS searched for HSCPs as it has a direct 
impact on some of the choices that we make in this paper.
We will then introduce a number of simplified models that contain one
or two HSCPs in the final state. These models correspond to the simplest topologies and
appear in several BSM theories.
The models are summarized in Tab.~\ref{tab:defModels1} (two HSCPs in the final
state) and Tab.~\ref{tab:defModels2} (one HSCP and one neutral BSM
particle in the final state).
For computing the efficiencies as a function of the topologies and the
BSM masses appearing in the cascade decays, we must choose a specific BSM model for
production and decay of the $\mathbb{Z}_2$-odd particles. Here we use the
supersymmetric (SUSY) simplified models, which are listed in the last column of
Tabs.~\ref{tab:defModels1} and~\ref{tab:defModels2}. In these, the HSCP
is either the lightest chargino, $\s\chi^{\pm}_1$, or the lighter stau,
$\stau_1$. As part of the Simplified Model approximations, we assume that
the efficiencies are weakly dependent on the spin of the HSCP\@.

\renewcommand{\arraystretch}{1.2}
\begin{table}[!ht]
\begin{center}
\begin{tabular}{lllr}
\toprule
\multicolumn{1}{l}{name} &\multicolumn{1}{c}{\!\!\!\!\!\!\!\!diagram} & \multicolumn{1}{l}{parameters} & SUSY topology\\
\midrule
$\mathcal{M}1$ &
\setlength{\unitlength}{1\textwidth}
\begin{picture}(0.3,0.1)
 \put(-0.015,-0.08){
  \put(0.0,0.025){\includegraphics[width=0.16\textwidth]{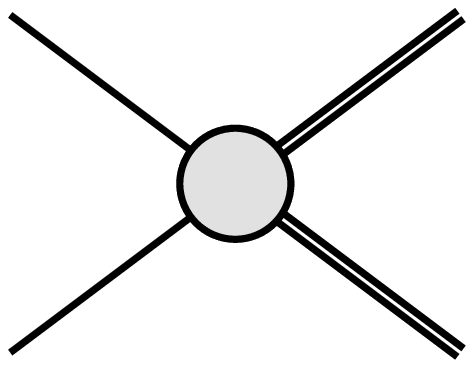}}
  \put(0.17,0.14){\footnotesize $m_{\text{HSCP}}$}
  \put(0.17,0.02){\footnotesize $m_{\text{HSCP}}$}
   }
\end{picture} & $m_{\text{HSCP}}$ & $pp\to \charg \charg$ \\
$\mathcal{M}3$ &
\setlength{\unitlength}{1\textwidth}
\begin{picture}(0.3,0.185)
 \put(-0.015,-0.12){
  \put(0.0,0.025){\includegraphics[width=0.22\textwidth]{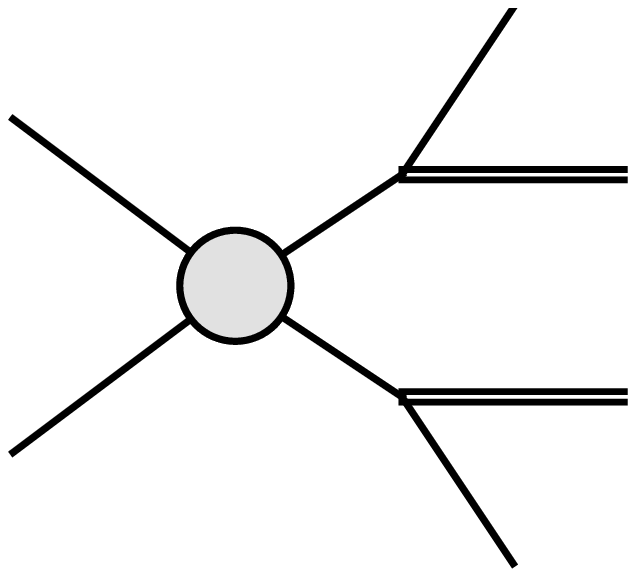}}
  \put(0.125,0.138){\footnotesize $m_{\text{prod}}$}
  \put(0.125,0.104){\footnotesize $m_{\text{prod}}$}
  \put(0.23,0.164){\footnotesize $m_{\text{HSCP}}$}
  \put(0.23,0.08){\footnotesize $m_{\text{HSCP}}$}
   }
\end{picture} & $m_{\text{HSCP}},m_{\text{prod}}$ & $pp\to \sq\sq\to \charg \charg$ \\
$\mathcal{M}5$ &
\setlength{\unitlength}{1\textwidth}
\begin{picture}(0.3,0.2)
 \put(-0.015,-0.12){ 
  \put(0.0,0.025){\includegraphics[width=0.295\textwidth]{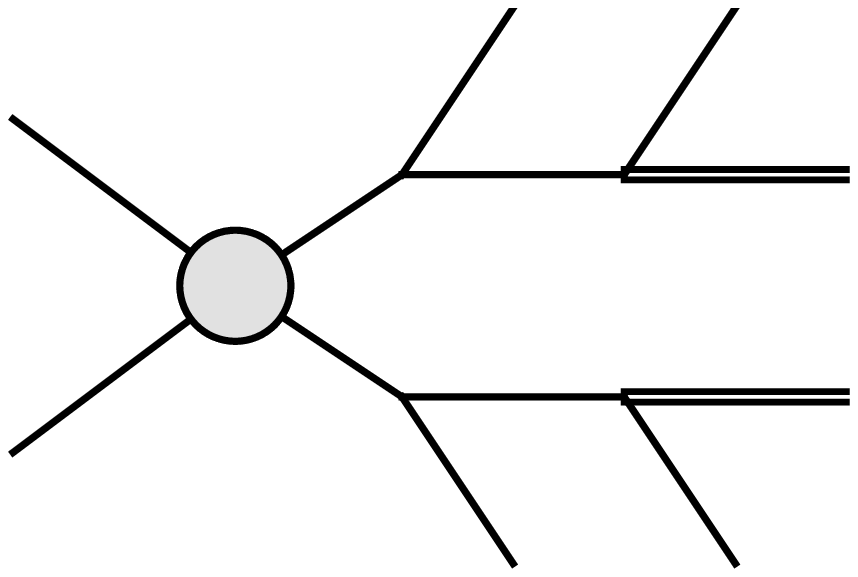}}
  \put(0.125,0.138){\footnotesize $m_{\text{prod}}$}
  \put(0.125,0.104){\footnotesize $m_{\text{prod}}$}
  \put(0.17,0.18){\footnotesize $m_{\text{int}}$}
  \put(0.17,0.066){\footnotesize $m_{\text{int}}$}
  \put(0.25,0.18){\footnotesize $m_{\text{HSCP}}$}
  \put(0.25,0.066){\footnotesize $m_{\text{HSCP}}$}
   }
\end{picture} & $m_{\text{HSCP}},m_{\text{int}},m_{\text{prod}}$ & $pp\to \sq\sq\to \neu \neu\to\stau_1\stau_1$ \\
$\mathcal{M}7$ &
\setlength{\unitlength}{1\textwidth}
\begin{picture}(0.3,0.2)
 \put(-0.015,-0.12){
  \put(0.0,0.025){\includegraphics[width=0.295\textwidth]{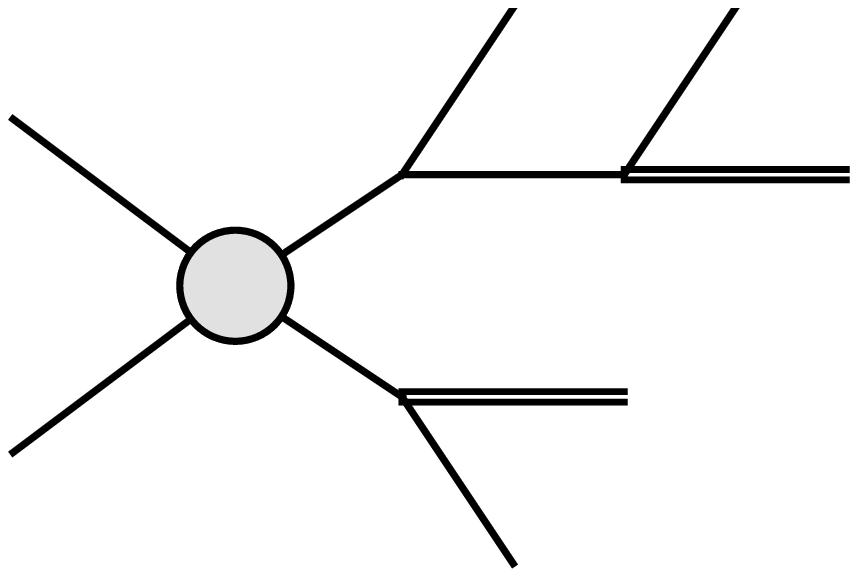}}
  \put(0.125,0.138){\footnotesize $m_{\text{prod}}$}
  \put(0.125,0.104){\footnotesize $m_{\text{prod}}$}
  \put(0.167,0.179){\footnotesize $m_{\text{int}}$}
  \put(0.17,0.066){\footnotesize $m_{\text{HSCP}}$}
  \put(0.25,0.18){\footnotesize $m_{\text{HSCP}}$}
   }
\end{picture} & $m_{\text{HSCP}},m_{\text{int}},m_{\text{prod}}$ & $pp\to \neu\s\chi^\pm_2\to \stau_1 (\charg\to\stau_1)$ \\
$\mathcal{M}8$ &
\setlength{\unitlength}{1\textwidth}
\begin{picture}(0.3,0.2)
 \put(-0.015,-0.12){ 
  \put(0.0,0.025){\includegraphics[width=0.22\textwidth]{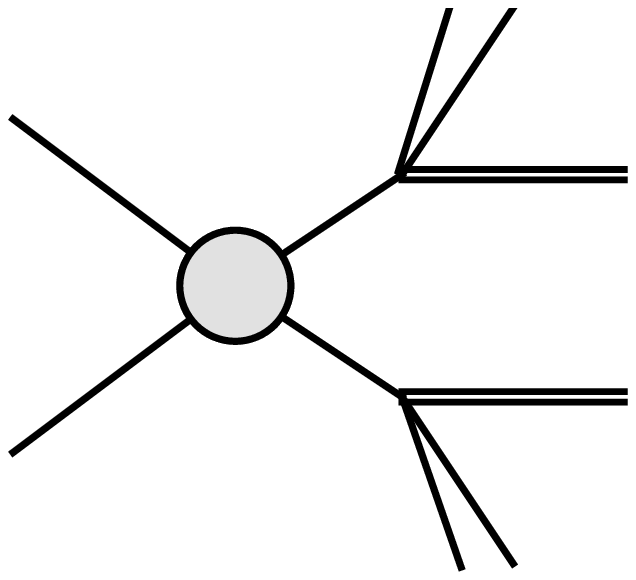}}
  \put(0.125,0.138){\footnotesize $m_{\text{prod}}$}
  \put(0.125,0.104){\footnotesize $m_{\text{prod}}$}
  \put(0.23,0.164){\footnotesize $m_{\text{HSCP}}$}
  \put(0.23,0.08){\footnotesize $m_{\text{HSCP}}$}
   }
\end{picture} & $m_{\text{HSCP}},m_{\text{prod}}$ & $pp\to \sq\sq\to \stau_1\stau_1$ \\
& & &  \\
& & &  \\
& & &  \\
\bottomrule
\end{tabular}
\end{center}
\caption{Definitions of the simplified models with two HSCPs used in this study. 
In the diagrams single solid lines denote SM
particles or intermediate BSM particles, double solid lines denote the HSCP\@. }
\label{tab:defModels1}
\end{table}

\renewcommand{\arraystretch}{1.2}
\begin{table}[!t]
\begin{center}
\begin{tabular}{lclr}
\toprule
\multicolumn{1}{l}{name} &\multicolumn{1}{c}{\!\!\!\!\!\!\!\!diagram} & \multicolumn{1}{l}{parameters} & SUSY topology\\
\midrule
$\mathcal{M}2$ &
\setlength{\unitlength}{1\textwidth}
\begin{picture}(0.3,0.1)
 \put(-0.015,-0.08){ 
  \put(0.0,0.025){\includegraphics[width=0.16\textwidth]{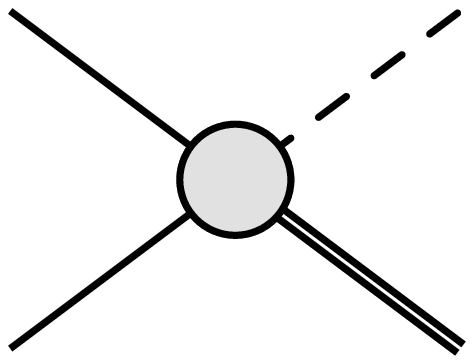}}
  \put(0.17,0.14){\footnotesize $m_{\text{inv}}$}
  \put(0.17,0.02){\footnotesize $m_{\text{HSCP}}$}
   }
\end{picture} & $m_{\text{HSCP}} = m_{\text{inv}}$ & $pp\to  \charg \neu$ \\
$\mathcal{M}4$ &
\setlength{\unitlength}{1\textwidth}
\begin{picture}(0.3,0.2)
 \put(-0.015,-0.12){
  \put(0.0,0.025){\includegraphics[width=0.22\textwidth]{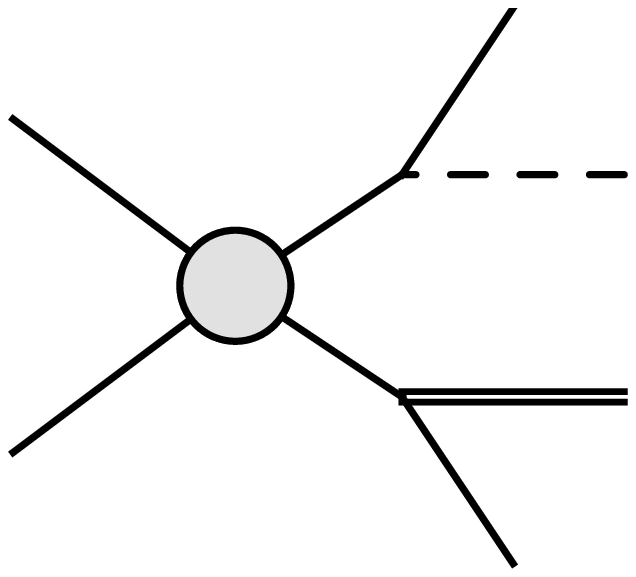}}
  \put(0.125,0.138){\footnotesize $m_{\text{prod}}$}
  \put(0.125,0.104){\footnotesize $m_{\text{prod}}$}
  \put(0.23,0.08){\footnotesize $m_{\text{HSCP}}$}
   }
\end{picture} & $m_{\text{HSCP}},m_{\text{prod}}$ & $pp\to \sq\sq\to \charg \neu$ \\
$\mathcal{M}6$ &
\setlength{\unitlength}{1\textwidth}
\begin{picture}(0.3,0.2)
 \put(-0.015,-0.12){ 
  \put(0.0,0.025){\includegraphics[width=0.295\textwidth]{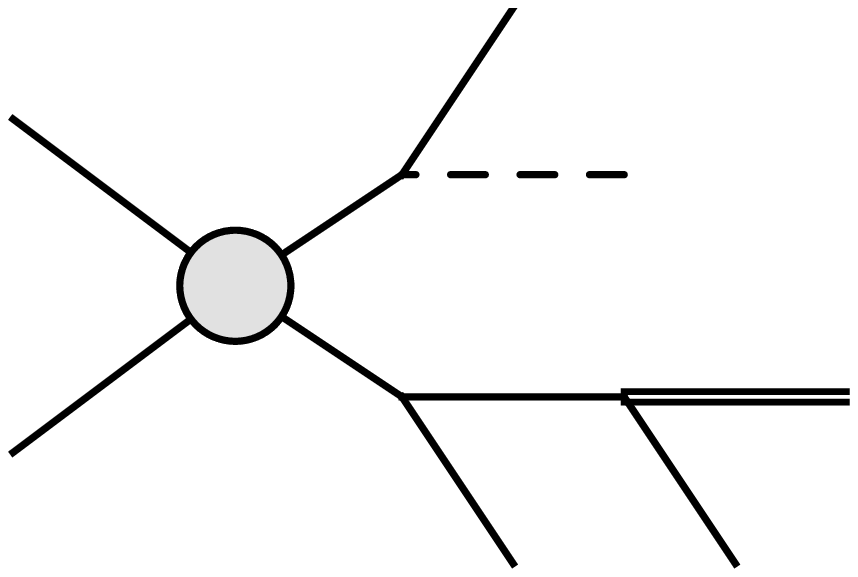}}
  \put(0.125,0.138){\footnotesize $m_{\text{prod}}$}
  \put(0.125,0.104){\footnotesize $m_{\text{prod}}$}
  \put(0.17,0.066){\footnotesize $m_{\text{int}}$}
  \put(0.25,0.066){\footnotesize $m_{\text{HSCP}}$}
   }
\end{picture} & $m_{\text{HSCP}},m_{\text{int}},m_{\text{prod}}$ & $pp\to \sq\sq\to \neu (\neu\to\stau_1)$ \\
& & &  \\
& & &  \\
& & &  \\
\bottomrule
\end{tabular}
\end{center}
\caption{Definitions of the simplified models with one HSCP used in this study. 
In the diagrams single solid lines denote SM
particles or intermediate BSM particles, double solid lines denote the HSCP and dashed lines denote invisible
particles (or an invisible branch, see Fig.~\ref{fig:invmapping}). }
\label{tab:defModels2}
\end{table}

\subsection{Overview of the CMS Search for HSCP} \label{sec:CMSOverview}

Heavy stable charged particles are highly penetrating particles that are expected to 
cross the entire CMS detector and reach the muon system\footnote{The behavior of 
color-charged HSCPs is more complex~\cite{CMS1305.0491}, but we focus only on 
the case of lepton-like HSCPs in this paper.}, and are therefore experimentally 
reconstructed and identified as muon particles.
However, because of their large mass and the \emph{limited} energy available in LHC 
collisions, they will be travelling through the detector with a velocity ($\beta$) significantly 
slower than the speed-of-light.
Consequently, they will have an anomalously high ionization energy loss ($\D E/\D x$) 
and a longer time-of-flight (TOF) than relativistic standard model particles.
In the CMS search for HSCP~\cite{CMS1305.0491}, the CMS silicon tracker is used to 
measure the particle $\D E/\D x$, while the CMS muon system is used to measure the 
particle's TOF\@. The events are mostly selected online by a muon ($\pt>45\,\GEV$) 
trigger. However, the trigger becomes inefficient when the particle velocity is too low 
($\beta<0.45$) due to the too long delay ($>25\,\text{ns}$) for the particle to reach the 
muon system causing a mismatch between the muon system information and the inner 
tracker information.

While there is no real standard model background to this search, instrumental 
backgrounds due to the mis-measurement of either $\D E/\D x$ or TOF is not negligible.
To predict the amount of backgrounds in the signal region CMS exploits the fact that the 
$\D E/\D x$ and TOF measurements are uncorrelated for backgrounds.
The track $\D E/\D x$ and momentum variables are used to reconstruct the particle mass 
and further discriminate the HSCP signal from mis-reconstruction background peaking 
at low values of the reconstructed mass.
Although the mass threshold used in~\cite{CMS1305.0491} is continuous, the required 
inputs for the reinterpretation of these results~\cite{Khachatryan:2015lla} are only 
provided in 100\,GeV steps. Below we use the results presented in
Ref.~\cite{Khachatryan:2015lla} to compute the signal efficiencies for the
simplified models introduced in Tables~\ref{tab:defModels1} and~\ref{tab:defModels2}.

\subsection{Computation of signal efficiencies} \label{sec:effGen}

In order to compute the efficiencies for the simplified models, we
perform a Monte Carlo simulation of the signal at the 8\,TeV LHC\@.
For each topology listed in Tabs.~\ref{tab:defModels1} and~\ref{tab:defModels2} 
we scan over the respective BSM masses (listed in the
third column) and generate 30\,k events for each set of masses. For the event
generation we use \textsc{MadGraph}~5~\cite{Alwall:2011uj} to generate
parton level events and then \textsc{Pythia}~6~\cite{Sjostrand:2006za}
to perform the decays, as well as showering and hadronization.
{\it No detector simulation is performed}, since we follow the
fast simulation procedure defined in Ref.~\cite{Khachatryan:2015lla},
where signal acceptances for HSCP candidates are provided as a function
of the HSCP's kinematics. In order to identify HSCP candidates in each event we
must first apply the following isolation criteria:
\begin{equation}
\left( \sum_{j}^{\stackrel{\text{charged particles}}{\Delta R<0.3}} \pt^{j}
\right)  < 50\GEV\;\;\; \text{ and }\;\;\; \left(
\sum_{j}^{\stackrel{\text{visible particles}}{\Delta R<0.3}} \frac{E^j}{
|\vec{p}|} \right)  < 0.3\,,
\label{eq:GenTkIso}
\end{equation}
where the first (second) sum includes all the charged (visible) particles
in a cone of $\Delta R=\sqrt{\Delta\eta^2+\Delta \phi^2}<0.3$ around the
direction of the long-lived particle, $\pt^{j}$ denotes their transverse
momenta, $E^j$ their energy and $|\vec{p}|$ is the magnitude of the long-lived
particle's three-momentum.
In both sums the long-lived particle candidate itself is not included.
As muons release very little energy in the calorimeters
they are not con\-si\-dered as visible particles.
The purpose of these isolation requirements is to mimic the event selection 
used in the CMS analysis~\cite{CMS1305.0491}.
Long-lived particles failing any of these isolation requirements are not
considered as HSCP candidates.

Once the HSCP candidates are identified, we can compute the signal
efficiencies using the acceptances provided in Ref.~\cite{Khachatryan:2015lla}.
These acceptances are given as probabilities for the candidate to pass the on-
and off-line selection criteria ($P_\text{on}$ and $P_\text{off}$)  and
depend on the  candidate's pseudo-rapidity $\eta$, transverse momentum $\pt$ and
velocity $\beta$. The final signal efficiency ($\epsilon$) is then given by:
\begin{equation}
\label{eq:Technique}
\epsilon = \frac{1}{N} \sum_{i}^{N}
P_{\text{on}}\left(\vec{k}_i\right) \times
P_{\text{off}}\left(\vec{k}_i\right)\,,
\end{equation}
where $P_{\text{on}}$ ($P_{\text{off}}$) is the on-line (off-line)
probability for each event, the sum runs over all generated events, $N$, and
$\vec{k}_i=(\eta_i,\pt_i,\beta_i)$ contains the kinematic properties for the
HSCP candidate in the $i$th event.
For events containing two HSCP candidates, the above probabilities must be
replaced by~\cite{Khachatryan:2015lla}
\begin{equation}
\label{eq:EventAcceptance}
P^{(2)}_{\text{on}/\text{off}}(\vec{k}^1_i, \vec{k}^2_i) 
= P_{\text{on}/\text{off}}(\vec{k}^1_i)  + P_{\text{on}/\text{off}}(\vec{k}^2_i) 
- P_{\text{on}/\text{off}}(\vec{k}^1_i)  P_{\text{on}/\text{off}}(\vec{k}^2_i)  \,,
\end{equation}
where $\vec{k}_i^{1,2}$ are the kinematical vectors of the HSCPs.
There are two main effects governing $P_{\text{on}/\text{off}}$.
On the one hand, the velocity $\beta$ should considerably deviate from 1 in
order to allow for a discrimination against muons. Hence, for $\beta\to1$
the acceptance goes to zero. On the other hand for 
too small $\beta$ ($\beta \lesssim 0.45$) the particle may not be assigned to
the right bunch crossing anymore. In this case, the trigger efficiencies
(online selection) go down very drastically.

The CMS analysis also requires a minimum reconstructed mass
($m_\text{rec}$) for the candidate. For the fast simulation method used
here, the collaboration provides the $P_{\text{on}/\text{off}}$ probabilities
for four distinct mass cuts, which we consider as four different signal
regions:
\begin{eqnarray*}
&\text{SR}_{0}: m_\text{rec} > 0 \GEV,\;\; \text{SR}_{100}: m_\text{rec} > 100
\GEV,\\
& \text{SR}_{200}: m_\text{rec} > 200 \GEV\;\; \text{ and }\;\;
\text{SR}_{300}: m_\text{rec} > 300 \GEV.
\end{eqnarray*}
Due to detector resolution effects, the reconstructed mass is typically 
$m_\text{rec} \simeq 0.6m_\text{HSCP} $~\cite{Khachatryan:2015lla}
and the above requirements must be
translated to the real HSCP mass.
Therefore, when computing the efficiencies for each signal region, we take
$\epsilon = 0$, if $m_\text{HSCP} < 166\,$GeV, 334\,GeV and 500\,GeV for the
signal regions SR$_{100}$, SR$_{200}$ and SR$_{300}$, respectively.

\subsection{Validation} \label{sec:validate}

\begin{figure}[!h]
\centering
\setlength{\unitlength}{1\textwidth}
\begin{picture}(1,0.47)
 \put(-0.015,-0.0){ 
  \put(0.07,0.14){\includegraphics[width=0.42\textwidth]{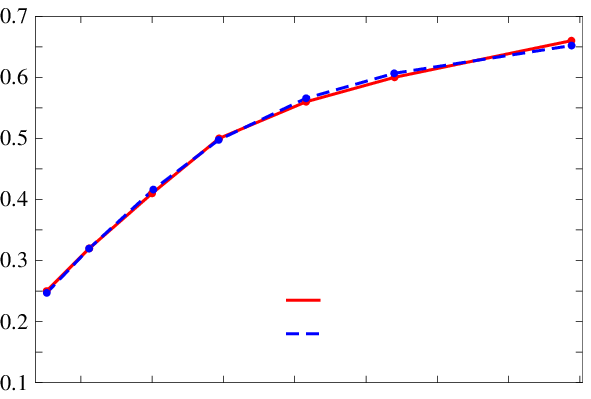}}
  \put(0.03,0.285){\rotatebox{90}{$\epsilon$}}
  \put(0.31,0.214){\rotatebox{0}{\scriptsize  \color{black} CMS-EXO-13-006 }}
  \put(0.31,0.19){\rotatebox{0}{\scriptsize \color{black} Our simulation
  }} \put(0.062,0.03){\includegraphics[width=0.4335\textwidth]{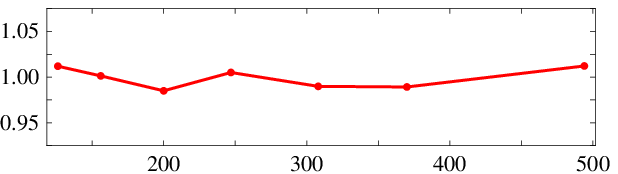}}
  \put(0.03,0.075){\rotatebox{90}{\footnotesize 
  }}
  \put(0.03,0.075){\rotatebox{90}{\footnotesize Ratio}}
  \put(0.25,0.0){\footnotesize $m_{\stau_1} \,[\text{GeV}]$}
  }
 \put(0.495,-0.0){ 
  \put(0.059,0.14){\includegraphics[width=0.431\textwidth]{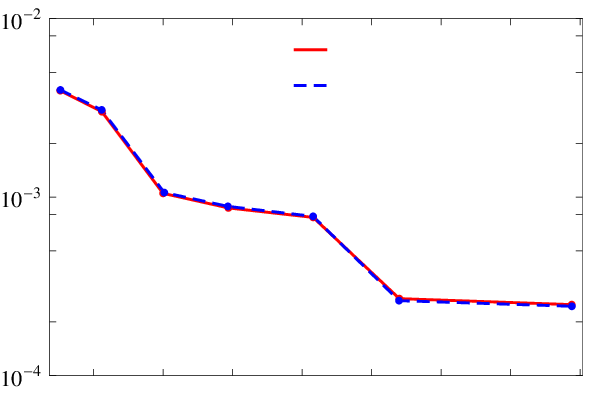}}
  \put(0.027,0.24){\rotatebox{90}{\footnotesize $\sigma_\text{limit}\;[\text{pb}]$}}
  \put(0.027,0.075){\rotatebox{90}{\footnotesize Ratio}}
  \put(0.31,0.394){\rotatebox{0}{\scriptsize  \color{black} CMS-EXO-13-006}}
  \put(0.31,0.367){\rotatebox{0}{\scriptsize \color{black} Our
  simulation }} \put(0.062,0.03){\includegraphics[width=0.4335\textwidth]{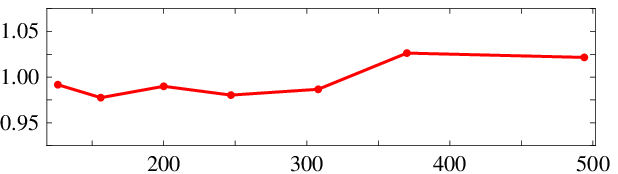}}

  \put(0.25,0.0){\footnotesize $m_{\stau_1} \,[\text{GeV}]$}
  }
\end{picture}
\caption{
Signal efficiency $\epsilon$ (left panel) and 95\% CL cross section upper limit
(right panel) for the GMSB model as the function of the stau mass. We compare the CMS analysis 
(CMS-EXO-13-006~\cite{Khachatryan:2015lla})
from the full detector simulation (red solid
lines) with our implementation of the analysis described in Sec.~\ref{sec:effGen} (blue dashed lines).
In the lower frames we show the respective ratios
$\epsilon^\text{CMS}/\epsilon^\text{Our}$,
$\sigma^\text{CMS}_\text{limit}/\sigma_\text{limit}^\text{Our}$.
}
\label{fig:gmsbComp}
\end{figure}

In order to validate the procedure described in Sec.~\ref{sec:effGen},
we compute the efficiencies for the gauge-mediated supersymmetry breaking (GMSB)
models considered by the CMS collaboration in Ref.~\cite{Khachatryan:2015lla}.
These models have a gravitino LSP and a long-lived stau as the NLSP\@.
As a result, for collider purposes, all the sparticles cascade decay to the
lightest stau, which is the HSCP candidate.  
We simulated the signal with \textsc{Pythia}~6 and analyzed the generated
events as described in Sec.~\ref{sec:effGen}. The results obtained for the
inclusive production of staus are shown in Fig.~\ref{fig:gmsbComp}, where we
also show the corresponding efficiencies obtained by the CMS collaboration.
As in Ref.~\cite{Khachatryan:2015lla}, we choose $\text{SR}_{0}$ for 
$m_\text{HSCP} < 166\,$GeV, $\text{SR}_{100}$ for $166\,\GEV<m_\text{HSCP}<
334\,\GEV$ and $\text{SR}_{200}$ for higher masses.
Our efficiencies agree within 3\% with the ones obtained by CMS,
where the differences are likely due to Monte Carlo statistical uncertainties.
Therefore our procedure for computing the signal efficiencies
reproduce very well the experimental results and can be used to produce
the efficiency maps for the simplified models listed in Tabs.~\ref{tab:defModels1} 
and~\ref{tab:defModels2}. In Fig.~\ref{fig:gmsbComp} we also reproduce the 95\% CL
limits on the inclusive production cross sections, which again agree very well (within
$\sim 3\%$) with the ones obtained by CMS from Ref.~\cite{Khachatryan:2015lla}. 
Note that this limit is based on the discrete mass cuts on $m_\text{rec}$ mentioned
above, the full CMS analysis allows for a event-based
mass cut, resulting in somewhat stronger constraints for some values of the
HSCP mass.

\subsection{Results for the efficiency maps} \label{sec:effRes}

\begin{figure}[!t]
\centering
\setlength{\unitlength}{1\textwidth}
\begin{picture}(0.55,0.36)
 \put(0.0,0.0){ 
  \put(0.07,0.03){\includegraphics[width=0.42\textwidth]{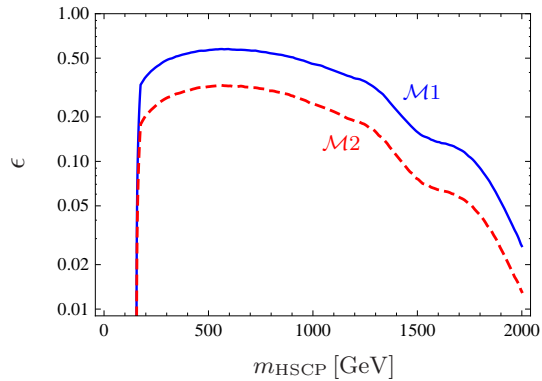}}
  \put(0.24,0.0){\footnotesize $m_\text{HSCP} \,[\text{GeV}]$}
  \put(0.03,0.17){\rotatebox{90}{ $\epsilon$}}
  \put(0.37,0.24){\rotatebox{0}{\scriptsize  \color{blue} $\mathcal{M} 1$}}
  \put(0.3,0.195){\rotatebox{0}{\scriptsize \color{red} $\mathcal{M} 2$ }}
  }
\end{picture}
\caption{
Signal efficiency $\epsilon$ as a function of the mass of the HSCP for model
$\mathcal{M} 1$ (direct production of two HSCPs, blue solid curve) and model
$\mathcal{M} 2$ (direct production of one HSCP and one invisible particle, red dashed curve).
}
\label{fig:M1M2eff}
\end{figure}

\begin{figure}[!t]
\centering
\setlength{\unitlength}{1\textwidth}
\begin{picture}(1,0.395)
 \put(-0.02,-0.005){ 
  \put(0.03,0.00){\includegraphics[width=0.5\textwidth]{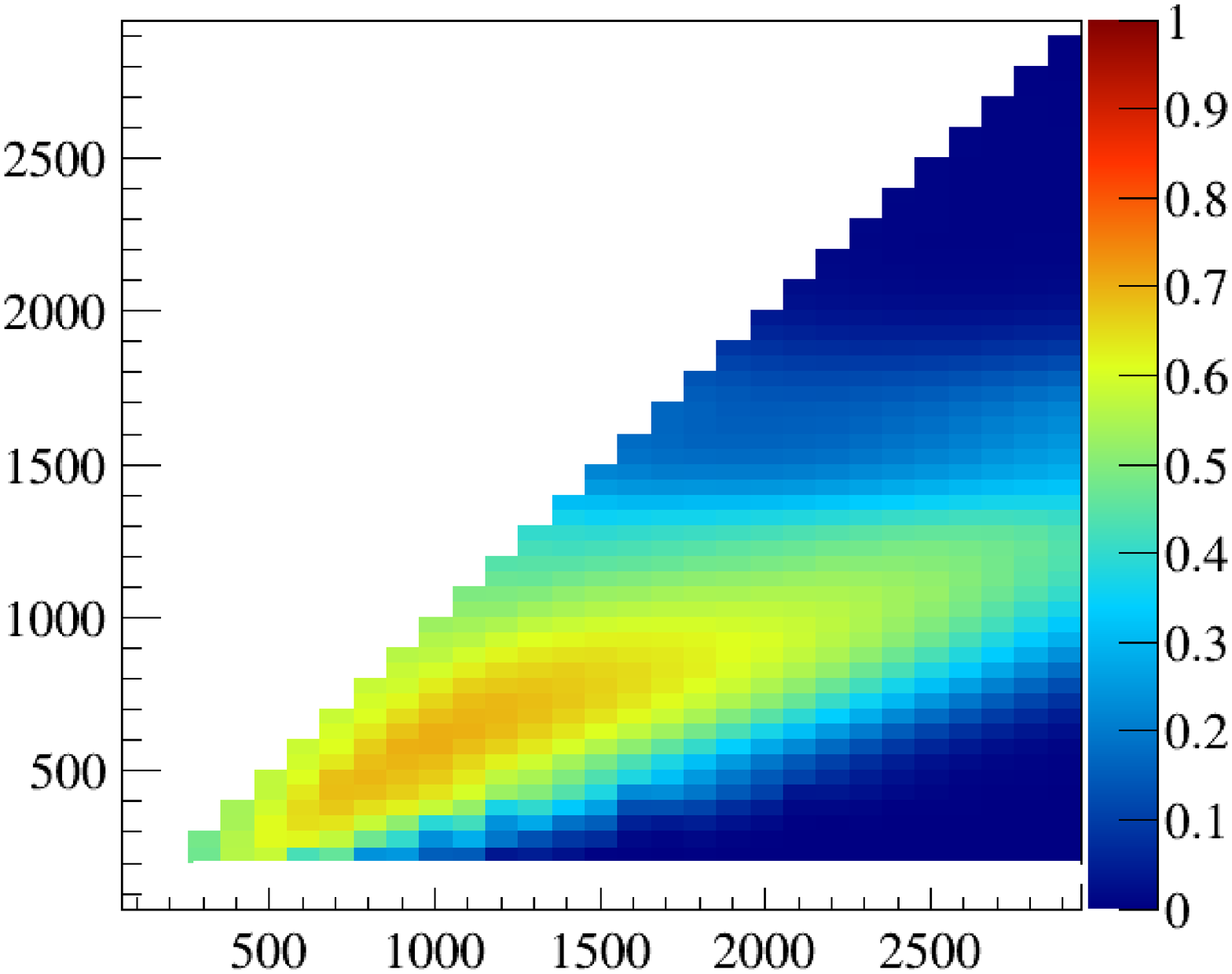}}
  \put(0.04,0.144){\rotatebox{90}{\footnotesize $m_\text{HSCP}\,[\text{GeV}]$}}
  \put(0.23,0.0){\footnotesize $m_\text{prod} \,[\text{GeV}]$}
  \put(0.497,0.215){\rotatebox{-90}{ $\epsilon$}}
  \put(0.14,0.315){\rotatebox{0}{\scriptsize  $\mathcal{M} 3$ }}
  }
 \put(0.49,-0.005){ 
  \put(0.03,0.00){\includegraphics[width=0.5\textwidth]{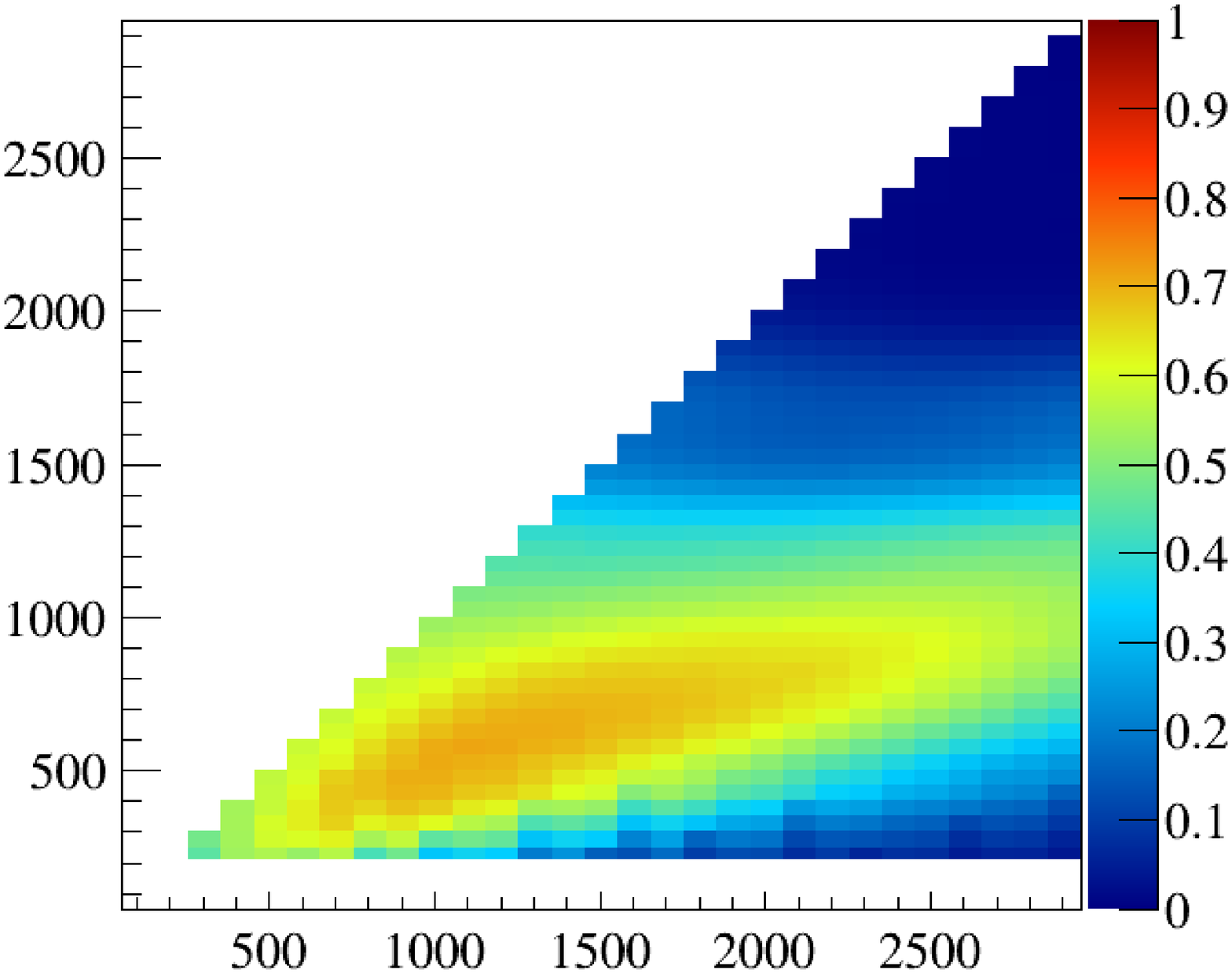}}
  \put(0.14,0.315){\rotatebox{0}{\scriptsize  $\mathcal{M} 8$ }}
  \put(0.04,0.144){\rotatebox{90}{\footnotesize $m_\text{HSCP}\,[\text{GeV}]$}}
  \put(0.23,0.0){\footnotesize $m_\text{prod} \,[\text{GeV}]$}
  \put(0.497,0.215){\rotatebox{-90}{ $\epsilon$}}
}
\end{picture}
\caption{
Signal efficiencies for model
$\mathcal{M} 3$ (left panel) and model
$\mathcal{M} 8$ (right panel).
}
\label{fig:M3M8eff}
\end{figure}

Our procedure to compute the efficiency maps using the method outlined in
Sec.~\ref{sec:effGen} is as follows. For the eight models listed in 
Tabs.~\ref{tab:defModels1} and \ref{tab:defModels2} we generate events and compute
the efficiencies in a wide range of sparticle masses between 50~GeV and 3~TeV,
varying the HSCP mass and all other masses listed in the tables in steps of
10~GeV and 50~GeV, respectively. In order to allow for a fast processing
within \textsc{SModelS} we then reduce the number of mass points in regions
of parameter space where the efficiencies do not vary considerably.
As an example, we show in Figs.~\ref{fig:M1M2eff} and \ref{fig:M3M8eff} the 
efficiencies for the SR$_{100}$ signal region ($m_\text{rec} > 100$~GeV or 
$m_\text{HSCP} > 166$~GeV) for the simplified models $\mathcal{M} 1$, 
$\mathcal{M} 2$ and $\mathcal{M} 3$, $\mathcal{M} 8$, respectively.

The signal efficiencies for the models with two HSCPs can be as high as
70\% and strongly depend on the HSCP mass. For direct production of two
HSCPs (simplified model $\mathcal{M} 1$), the signal efficiency stays above 20\%
for masses between 166~GeV and 1.4~TeV.
As discussed in Sec.~\ref{sec:effGen}, we take
$\epsilon = 0$ for $m_{\text{HSCP}} < 166$ GeV in the SR$_{100}$ signal region.
This is the reason for the sharp drop in the efficiencies in the light HSCP region,
seen in Fig.~\ref{fig:M1M2eff}.
For very large masses the particles are produced very close to threshold, with
extremely small $\beta$ ($\lesssim 0.5$) and the signal efficiency drops again.
This is a result of the low detection probabilities (in particular the
on-line probability) due to small trigger efficiencies for $\beta<0.5$.
In models  $\mathcal{M} 3$ and $\mathcal{M} 8$ the HSCPs are produced
in the decay of heavier particles and the efficiencies are largest 
in the intermediate mass range 500\,GeV to 1.2\,TeV. For large mass gaps
between the produced particle and the HSCP, the latter becomes
extremely boosted making the discrimination against the muon
background extremely hard. Hence, the signal efficiency decreases rapidly 
in this region. This decrease is less pronounced in model $\mathcal{M}8$, 
as the three-body decay leaves less energy to the HSCP\@.

As mentioned above, in the SMS framework
used here we neglect sub-leading effects due to the spin of the HSCP or any
other BSM particle. Nonetheless, we checked that, for direct production of the
HSCP, the differences in the efficiencies are smaller than 20\% when comparing
the $s$-channel production of spin 0, $1/2$ and 1 particles as well as the 
$t$-channel production of spin $1/2$ particles. For the production of BSM 
particles through longer decay chains we expect the differences to be
even smaller, as the kinematics of the HSCP are more strongly influenced
by the mass spectrum of the model. We hence expect that neglecting 
spin effects as well as the effects of the production channels ($s$-channel
versus $t$-channel) generates uncertainties of the order of 20\% or below.
Since these uncertainties are of the order of other theoretical uncertainties
(such as NLO corrections to the sparticles production cross sections), we
consider them acceptable.

\section{Using Simplified Models to Constrain Full Models} \label{sec:sms}

The efficiency maps described in Sec.~\ref{sec:effRes} allows us to compute 
the predicted signal cross section ($\sigma_\text{th}$) for a given simplified model
$\mathcal{M}_i$ in one of the four signal regions ($\text{SR}_j$):
\begin{equation}
\left(\sigma_\text{th}^{\mathcal{M}_i}\right)_{\text{SR}_j} =
\sigma^{\mathcal{M}_i} \times \epsilon^{\mathcal{M}_i}_{\text{SR}_j}\,,
\end{equation}
where $\sigma^{\mathcal{M}_i}$ is the cross section for
the simplified model and $\epsilon^{\mathcal{M}_i}_{\text{SR}_j}$
the respective efficiency for the signal region $\text{SR}_j$.
Comparing $\sigma_\text{th}$ with the experimental upper limit for the signal
cross section in the respective signal region ($\sigma_\text{UL}$), it is
possible to determine if the simplified model is excluded or not by the
experimental searches.
However, simplified models rarely match any model of
interest and their usefulness relies on the fact that, under some
approximations, it is possible to decompose a full model in terms of a coherent
sum of simplified models (see Ref.~\cite{Kraml:2013mwa} for details).
In this case, the full model signal cross section
($\sigma^\text{Full}_\text{th}$) to be confronted with $\sigma_\text{UL}$ is approximately
given by:
\begin{equation}
\left(\sigma^\text{Full}_\text{th}\right)_{\text{SR}_j} = \sum_{i}
\tilde{\sigma}^{\mathcal{M}_i} \times \epsilon^{\mathcal{M}_i}_{\text{SR}_j} \label{eq:sigfull}
\end{equation}
In the above expression $\tilde{\sigma}^{\mathcal{M}_i}$ is the corresponding
{\it weight} for the simplified model $\mathcal{M}_i$ in the full model.
These weights are computed by the decomposition procedure, which maps  the full
model into a sum of simplified models {\it topologies}. Since the decomposition
method used here follows closely the one used by \textsc{SModelS}~\cite{Kraml:2013mwa},
we only outline the main steps, focusing on the differences
required to treat long-lived particles.

First, using as input an SLHA card, all the widths ($\Gamma$), branching
ratios (BRs), masses and production cross sections ($\sigma$) of the
BSM states are defined for the input model.\footnote{In order to read the SLHA file, SModelS uses
the tools provided by the \textsc{PySLHA}~\cite{Buckley:2013jua} code.}
Second, for each of the particles appearing in the production cross sections, we
generate all possible cascade decays\footnote{Since the
total number of all possible cascade decays is typically of the order of
hundreds of thousands, we neglect all topologies which have
$\tilde{\sigma}^{\mathcal{M}_i} < \tilde{\sigma}_{\min}$. In the results
presented below we take $\tilde{\sigma}_{\min} = 5 \times 10^{-3} \,\text{fb}
\left(5 \times 10^{-4}\,\text{fb}\right)$ for the points with $m_\text{HSCP} \leq 400\,$GeV 
($m_\text{HSCP} > 400\,$GeV).}, using the
branching ratios and widths read from the SLHA card.
However, since now the input model may contain quasi-stable states, 
we must determine if the particle at the end
of the cascade decay is long-lived or not. More specifically, we must estimate
what is the probability for a BSM state to have a prompt decay or to decay
outside the detector. The fraction of particles which survive after traveling a
distance $l$ in the detector is given by:
\beq
f(l) = e^{-\Gamma l/\left(\gamma \beta\right)}
\eeq
where $\Gamma$ is the particle's width, $\gamma = 1/\sqrt{1-\beta^2}$ and
$\beta$ is the particle's boost.
Therefore, the probability for a particle to decay promptly is:
\beq
\mathcal{F}_\text{prompt} = 1-e^{-\Gamma l_{\text{inner}}/\left(\gamma \beta\right)},
\eeq
where $l_{\text{inner}}$ corresponds to the inner size of the
detector, for which all decays are seen as prompt. For our subsequent results we
take $l_{\text{inner}} = $ 10 mm.
On the other hand, the probability for a particle to decay {\it outside} the
detector is given by:
\beq
\mathcal{F}_\text{long} = e^{-\Gamma l_{\text{outer}}/\left(\gamma \beta\right)},
\eeq
where $l_{\text{outer}}$ corresponds to the detector size, which
we take to be 10\,m (for CMS).
Clearly the above probabilities are event-dependent, since they
depend on the boost of the unstable particle, through the factor $\gamma \beta$.
Nonetheless we can still conservatively estimate these. Since the
efficiencies for a long-lived particle to be identified as a charged track fall
sharply below $\beta \simeq 0.45$,
here we take $(\gamma \beta)_\text{outer} = 0.6$ (or $\beta \simeq
0.5$), which gives a mostly conservative estimate of $\mathcal{F}_\text{prompt}$.
On the other hand, for prompt decays we take $(\gamma \beta)_\text{inner} = 10$,
which corresponds to $\beta \simeq 0.995$. Notice that for most models, $\Gamma$
is such that the particle can be considered as decaying promptly or long-lived for a 
wide range of $\gamma\beta$ values. Therefore, for most cases, our results are only 
mildly dependent on our choice of $\gamma \beta$.

Once $\mathcal{F}_\text{prompt}$ and $\mathcal{F}_\text{long}$ are known for each 
state, during the decomposition each unstable particle (with a non-zero width) will
generate two possible topologies\footnote{The case of displaced vertices is not
considered in the present work.}:
\begin{itemize}
  \item one where the particle does not decay (it is considered as long-lived).
  In this case the topology weight will be proportional to the probability for
  the particle to decay outside the detector ($\mathcal{F}_\text{long}$);
  \item one where the particle decays promptly.
  In this case the topology weight will be proportional to the probability for
  the particle to decay inside the {\it inner} detector ($\mathcal{F}_\text{prompt}$).
\end{itemize}
Notice that in most cases we have
$\left(\mathcal{F}_\text{long},\mathcal{F}_\text{prompt}\right) \simeq (1,0)$ for
quasi-stable (or stable) particles or
$\left(\mathcal{F}_\text{long},\mathcal{F}_\text{prompt}\right) \simeq (0,1)$ for
unstable particles.

\begin{figure}[!t]
\centering
\setlength{\unitlength}{1\textwidth}
\includegraphics[width=\linewidth]{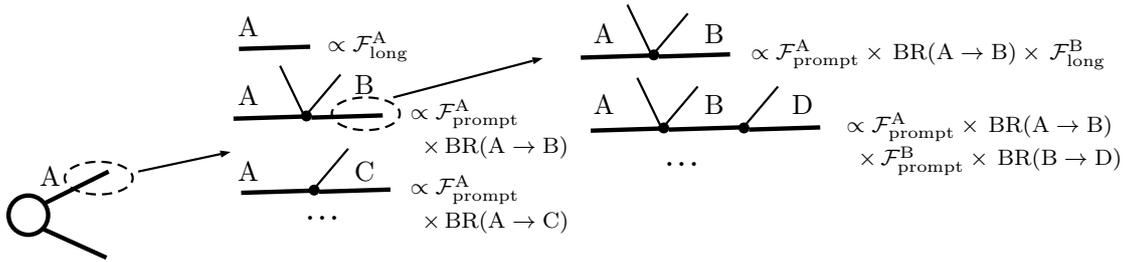}
\vspace{-5ex}
\caption{Decomposition procedure for scenarios containing quasi-stable
particles. Next to the topologies we show the factors contributing to the
topology weight. See text for details.}
\label{fig:decScheme}
\end{figure}

Using the information in the SLHA input and a modified version of the
package \textsc{SModelS}, which includes the computation of the
$\mathcal{F}_\text{prompt/long}$ probabilities, it is possible
to decompose full models into distinct simplified model topologies and
compute their weights.
As mentioned above, the decomposition procedure starts with the pair
production of BSM states (determined by the cross sections read from the SLHA
card) and generates all possible cascade decays for each particle produced.
For each step in the cascade decay, the topology weight is given by the
product of the production cross section ($\sigma_\text{prod}$), the BRs for the
decays appearing in the decay chain and the prompt decay/long-lived fractions,
$\mathcal{F}_\text{prompt/long}$:
\begin{equation}
\tilde{\sigma}^{\mathcal{M}_i} = \sigma_\text{prod} \times \left(\prod_i
\text{BR}_i \times \mathcal{F}_\text{prompt}^i \right) \times
\mathcal{F}_\text{long}^X \mathcal{F}_\text{long}^Y \,,
\end{equation}
where $\mathcal{F}_\text{long}^X \mathcal{F}_\text{long}^Y$ is the product of
the (non-decay) probabilities for the final states ($X,Y$) appearing in the cascade
decay chain.
This procedure is outlined in Fig.~\ref{fig:decScheme}. Since we only keep the
topologies with a final weight above a minimum value ($\tilde{\sigma}_{\min}$), 
the topologies containing very small probabilities ($\mathcal{F}_\text{prompt} \ll 1$ 
or $\mathcal{F}_\text{long} \ll 1$) are automatically discarded.
This procedure has the advantage of allowing us to probe scenarios where
more than one particle is (meta-)stable and to automatically determine which
states can be considered as long-lived or decaying promptly.
Furthermore, for models containing both neutral and charged (meta-)stable
particles, the above procedure will produce topologies with both HSCP and
missing energy (MET) signatures (or mixed MET-HSCP). Therefore it allows us to
simultaneously confront the corresponding model with both MET and HSCP
searches.

\begin{figure}[!h]
\centering
\setlength{\unitlength}{1\textwidth}
\begin{picture}(0.76,0.24)
 \put(0.0,0){ 
  \put(0.0,-0.011){\includegraphics[width=0.75\textwidth]{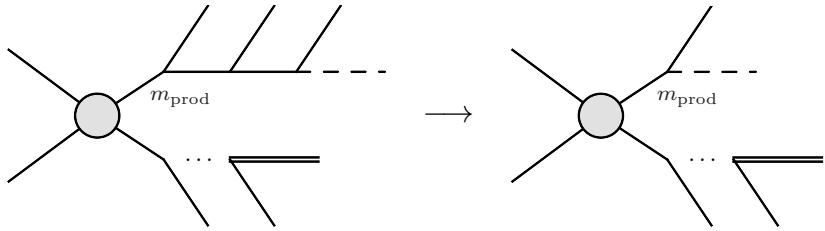}}
  \put(0.145,0.116){\scriptsize $m_{\text{prod}}$}
  \put(0.586,0.116){\scriptsize $m_{\text{prod}}$}
  }
\end{picture}
\caption{Models with one (arbitrarily long) decay chain terminating in an invisible particle 
are mapped onto the respective model with a single step decay in the
invisible (MET) branch.}
\label{fig:invmapping}
\end{figure}

Once the topology weights ($\tilde{\sigma}^{\mathcal{M}_i}$) are known,
through Eq.~\ref{eq:sigfull} it is possible to compute the full model signal
cross section for each of the signal regions considered by the experimental
searches. It is important to notice that, since we only computed efficiencies
for the simplified models appearing in Tabs.~\ref{tab:defModels1}
and~\ref{tab:defModels2}, the signal from any other topologies appearing during
the decomposition procedure are not included in the final signal. Although this
leads to conservative predictions, we will show below that, for the models
studied here we only underestimate the signal cross section by 20\% or less.
At this point it is also relevant to stress that the current public version of
\textsc{SModelS} does not include efficiency maps for the MET searches. For
these the topology weights cannot be summed up and the experimental upper
limits for the individual $\tilde{\sigma}^{\mathcal{M}_i}$ weights are used
instead (for more details see Ref.~\cite{Kraml:2013mwa}).
Since the decomposition procedure can produce topologies with one invisible and
one HSCP in the final states (such as the ones shown in
Tab.~\ref{tab:defModels2}), these can be constrained by both MET and HSCP
searches. However, since HSCP constraints are typically stronger,
the mixed topologies are considered as containing a single HSCP
and we only apply the constraints from HSCP searches.
Furthermore, since the HSCP efficiencies for these mixed topologies are almost
independent of the cascade decay ending in the invisible state, we can neglect
the kinematics of the MET branch and compress it to a single step decay, as shown in 
Fig.~\ref{fig:invmapping}. This allows us to use the efficiencies computed for
the simplified models $\mathcal{M}_4$ and $\mathcal{M}_6$ for
a wide range of HSCP-MET topologies and improve the coverage of our
efficiency maps.

Using the efficiency maps for the simplified models
$\mathcal{M}1$-$\mathcal{M}8$ computed in Sec.~\ref{sec:effRes} and the
decomposition procedure described above, we proceed to apply our modified
version of \textsc{SModelS} to a physical model of interest. As we will show,
both MET and HSCP searches can be relevant (although the former are typically
stronger) and allows us to impose strong constraints on the stau co-annihilation
region of the CMSSM.

\section{Application to the Lithium-7 Problem} \label{sec:appl}

In this section we apply the procedure outlined in Sec.~\ref{sec:sms}
to a full model containing long-lived particles. 
One interesting motivation for the existence of long-lived particles (in
cosmological scales) is the Lithium-7
problem~\cite{Spite:1982dd, Cyburt:2008kw} (for a recent review, see
Ref.~\cite{Cyburt:2015mya}).
Despite the enormous success of BBN, the Lithium
abundance inferred from the Cosmic Microwave Background and
BBN~\cite{Cyburt:2015mya}, 
\begin{equation}
\left(\frac{^7\text{Li}}{\text{H}}\right)_\text{theo} = \left(4.68 \pm 0.67\right) \times
10^{-10},
\end{equation}
is highly inconsistent with the experimentally measured Lithium
abundance~\cite{Sbordone:2010zi}:
\begin{equation}
\left(\frac{\text{Li}}{\text{H}}\right)_\text{exp} = \left(1.6 \pm 0.3\right) \times 10^{-10}\,.
\end{equation}
Although some of the proposed solutions to the above discrepancy
do not involve new physics (e.g. stellar depletion or inclusion of new nuclear
reactions), these are usually highly tuned or require modification of
nuclear rates well outside the expected uncertainties~\cite{Boyd:2010kj}.
A popular alternative is to invoke new physics during BBN, which could explain a
smaller Lithium production rate (or an annihilation of the original Lithium
abundance). A well studied scenario is supersymmetry with long-lived staus
($\stau$).
If $\stau$s are still present during BBN, they can form bound states with
nuclei (such as $^{7}$Li) and deplete the Lithium abundance, thus providing
a viable solution to the Lithium problem.
Since such solutions have been discussed at length in the 
literature~\cite{Jittoh:2007fr,Konishi:2013gda,Jedamzik:2004er,Jedamzik:2005dh,
Pospelov:2006sc,Cyburt:2006uv,Jedamzik:2007cp,Kusakabe:2007fu,Jittoh:2008eq,
Pospelov:2008ta,Jittoh:2010wh,Cyburt:2012kp,Kusakabe:2014moa}, 
here we concentrate on their features relevant for the LHC searches.

In this work we focus on the case of a neutralino LSP and consider the
CMSSM, closely following the discussion presented in Ref.~\cite{Konishi:2013gda}.
In order to cover the CMSSM parameter space we perform a Monte Carlo scan in 
the input parameters:
\begin{equation}
m_0, M_{1/2}, A_0\,,
\end{equation} 
where $m_0$ is the universal soft scalar mass, $M_{1/2}$ is the universal soft
gaugino mass and $A_0$ the trilinear soft term, all defined at the unification
scale, $M_\text{GUT} \simeq 2\times 10^{16}\GEV$. We take the supersymmetric
mass term $\mu$ to be positive ($\mu >0$), while we fix the ratio of the Higgs vacuum
expectation values to be $\tan\beta = 10$. We also limit our results to negative values 
of $A_0$, since these enhance the radiative corrections to the Higgs mass. Similar 
results would be obtained for $\mu<0$ and $A_0>0$ except in this case we would 
have a larger discrepancy between the predicted and measured values for the 
anomalous magnetic moment of the muon $(g-2)_\mu$. The supersymmetric spectrum 
is generated with \textsc{SPheno}~3.2.1~\cite{Porod:2011nf}, the
sparticle production cross sections are computed using \textsc{Pythia}~6
and \textsc{NLLfast}~\cite{Beenakker:1996ch,Beenakker:1997ut,Kulesza:2008jb,
Kulesza:2009kq,Beenakker:2009ha,Beenakker:2010nq,Beenakker:2011fu} and the 
neutralino relic abundance is computed with 
\textsc{micrOMEGAs}~3.0.24~\cite{Belanger:2013oya}.

As computed in Refs.~\cite{Jittoh:2007fr,Jittoh:2010wh}, in order to solve the
Lithium problem, the stau yield and life-time must satisfy:
\begin{equation}
Y^{0}_{\stau} \gtrsim 10^{-13} \mbox{ and } \tau_{\stau} \gtrsim 1-100s.
\end{equation}
The latter condition requires the stau to be nearly degenerate with the LSP, which we assume 
to be the lightest neutralino. In particular, the mass difference $\delta m = \mstau - \mne$ 
must be significantly smaller than the $\tau$ mass ($\delta m < 0.1$ GeV).
In this quasi-degenerate scenario, the stau abundance before its decay is
related to the neutralino relic abundance
by~\cite{Jittoh:2010wh}
\beq
Y^{0}_{\stau}\simeq\frac{Y_{\neu}}{2\left(1+\E^{\delta m /
T_{\text{f}}}\right)} \label{eq:ytau}\,,
\eeq
where $Y^0_{\stau}$ is the stau yield after freeze-out (and before
decay) and $Y_{\neu}$ is the final neutralino yield (after staus have
decayed), which can be obtained from its final relic abundance:
\begin{equation}
Y_{\neu} = \left(\frac{\omg}{2.741 \times 10^8}\right)
\left(\frac{\GEV}{\mne}\right). \label{eq:yneu}
\end{equation}
The neutralino freeze-out temperature, $T_\text{f}$, can be well approximated by $T_\text{f}
\simeq \mne/25$ for the parameter space considered below.

Before discussing the LHC constraints from MET and
HSCP searches, we first impose the following set
of minimal constraints to the CMSSM:

\begin{itemize}
  \item a neutralino LSP; \label{condLSP}
  \item $\delta m = \mstau - \mne < 0.1$ GeV;
  \item $Y_{\stau}^0 > 10^{-13}$;
  \item $120\GEV < m_h < 130\GEV$.\footnote{This loose interval on the Higgs
  mass window is due to the large theoretical uncertainties on the Higgs mass
  calculation in the MSSM. Furthermore, a more strict choice for the
  Higgs mass interval would not change our subsequent results.}
\end{itemize}
Although we keep points with $\omg > 0.1289$ (which violate
the {\it Planck}'s $3\sigma$ upper bound on the Dark Matter relic
abundance~\cite{Ade:2013zuv}), we will explicitly identify in our results
the region consistent with {\it Planck}.

Since the left-right mixing of staus is proportional 
to $A_\tau-\mu\tan\beta$, the stau mass not only depends on
the scalar mass parameter $m_0$ but also strongly on $A_0$ and $\tan\beta$.
On the other hand, the neutralino mass is mainly dependent on the
gaugino mass parameter $M_{1/2}$. Therefore, the
requirement $\mstau \simeq \mne$ introduces a
correlation between $M_{1/2}$ and $m_0, A_0, \tan\beta$.
In particular, as $m_0$ increases (for a fixed $M_{1/2}$ value), $A_0$ must
increase (in absolute value) in order to enhance the stau mixing and reduce its
mass, which must satisfy $\mstau \simeq \mne \propto M_{1/2}$.
Therefore, whilst scanning over $A_0$ and $M_{1/2}$ with flat priors, we limit
the scan over $m_0$ to a gaussian distribution around the value predicted by 
the linear relations found in \cite{Konishi:2013gda}. This dramatically increases 
the efficiency of obtaining points fulfilling the $\delta m < 0.1\GEV$ requirement.
We also checked that points outside the $2\sigma$-band of the gaussian
distribution never satisfy the mass splitting condition.

\subsection{Scan Results} \label{sec:results}

\begin{figure}[t]
\centering
\setlength{\unitlength}{1\textwidth}
\begin{picture}(1,0.74)
 \put(-0.02,0.38){ 
  \put(0.04,0.005){\includegraphics[width=0.47\textwidth]{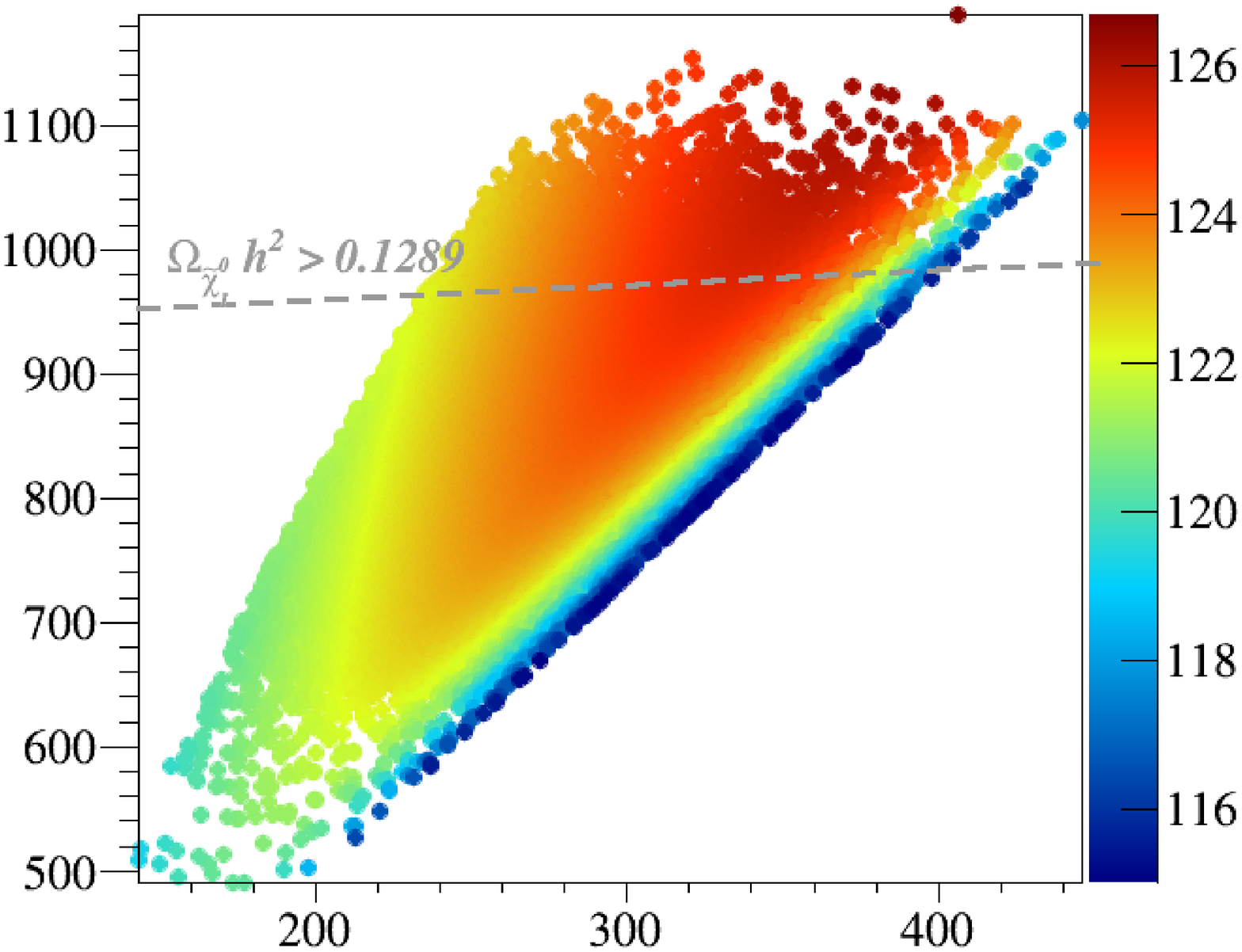}}
  \put(0.04,0.134){\rotatebox{90}{\footnotesize $M_{1/2}\,[\text{GeV}]$}}
  \put(0.226,0.0){\footnotesize $m_0 \,[\text{GeV}]$}
  \put(0.48,0.233){\rotatebox{-90}{\footnotesize $m_h \,[\text{GeV}]$}}
  }
 \put(0.48,0.38){ 
  \put(0.04,0.005){\includegraphics[width=0.47\textwidth]{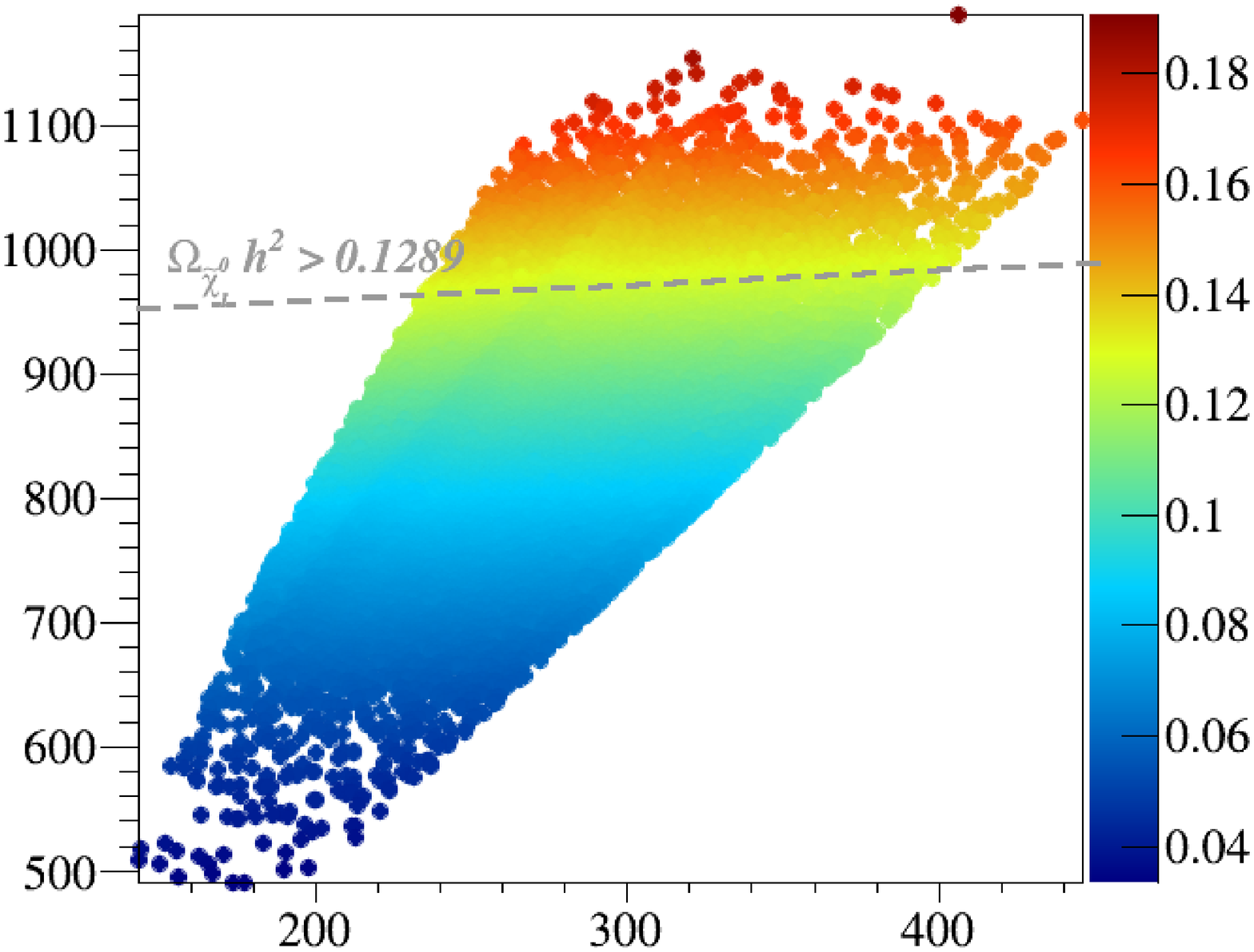}}
  \put(0.04,0.134){\rotatebox{90}{\footnotesize $M_{1/2}\,[\text{GeV}]$}}
  \put(0.226,0.0){\footnotesize $m_0 \,[\text{GeV}]$}
  \put(0.48,0.22){\rotatebox{-90}{\footnotesize $\omg$}}
}
 \put(-0.02,-0.0){ 
  \put(0.04,0.005){\includegraphics[width=0.47\textwidth]{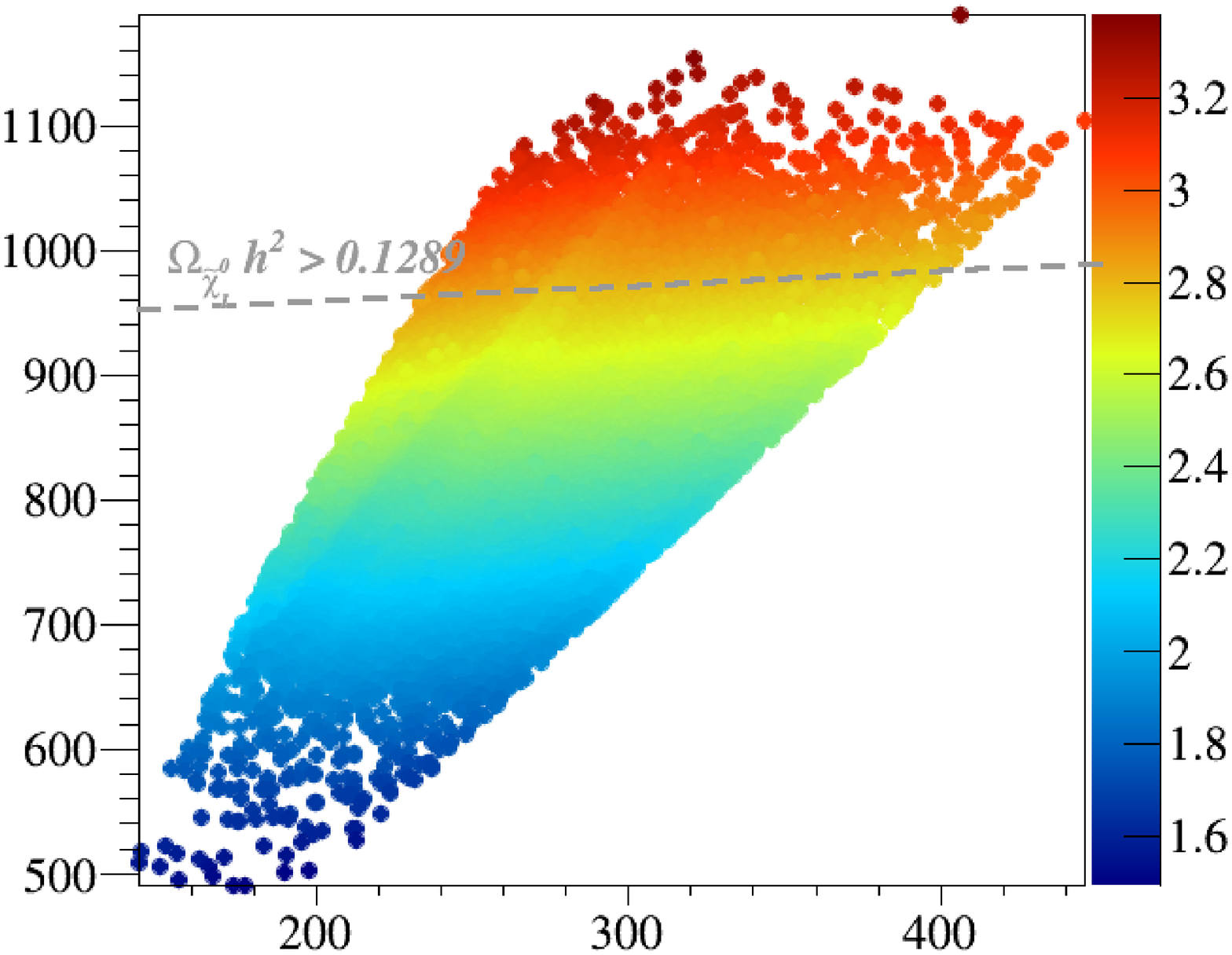}}
  \put(0.04,0.134){\rotatebox{90}{\footnotesize $M_{1/2}\,[\text{GeV}]$}}
  \put(0.226,0.0){\footnotesize $m_0 \,[\text{GeV}]$}
  \put(0.48,0.233){\rotatebox{-90}{\footnotesize $Y_{\stau}^0/10^{-13}$}}
  }
 \put(0.48,-0.0){ 
  \put(0.04,0.005){\includegraphics[width=0.47\textwidth]{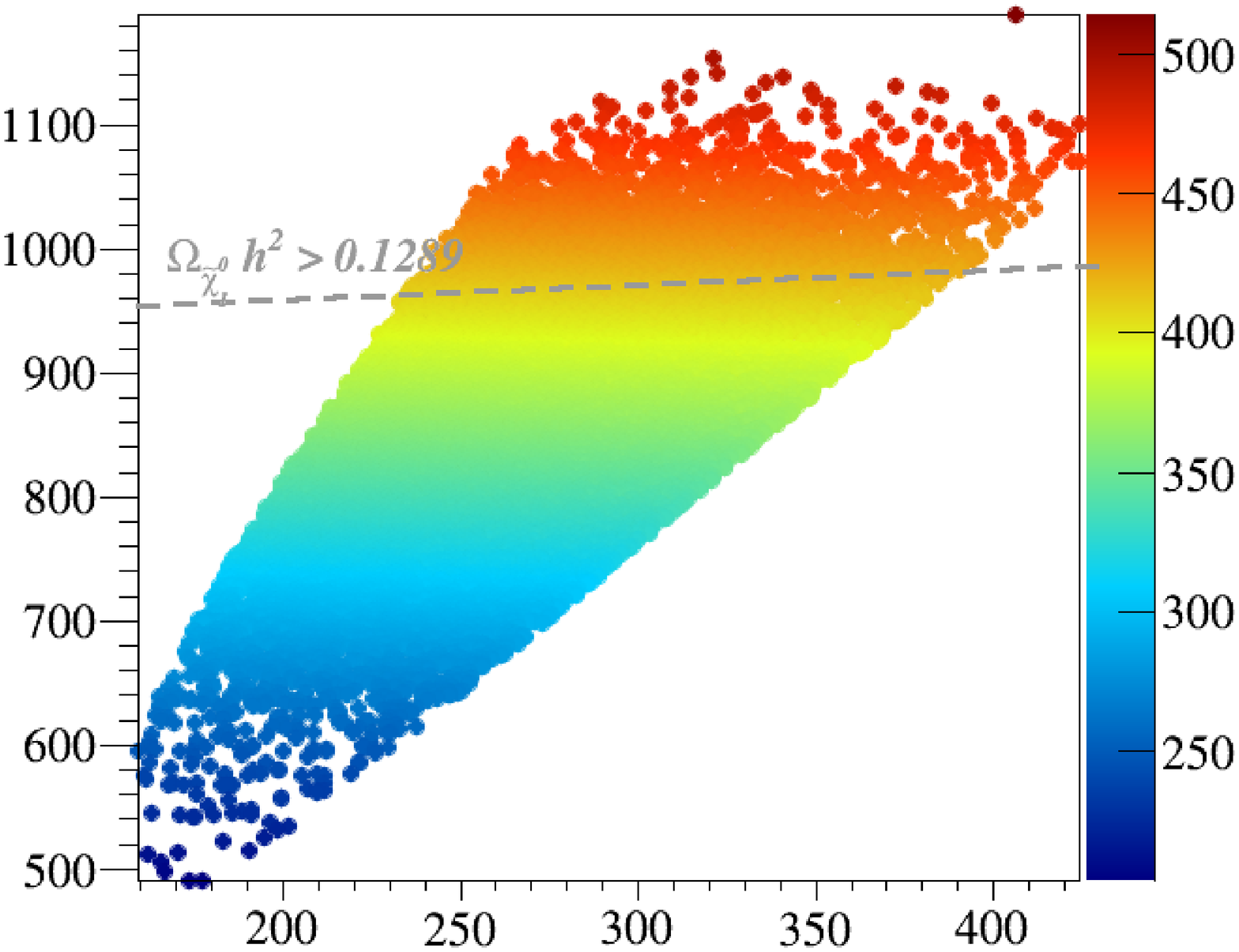}}
  \put(0.04,0.134){\rotatebox{90}{\footnotesize $M_{1/2}\,[\text{GeV}]$}}
  \put(0.226,0.0){\footnotesize $m_0 \,[\text{GeV}]$}
  \put(0.48,0.235){\rotatebox{-90}{\footnotesize $\mstau\,[\text{GeV}]$}}
}
\end{picture}
\caption{Values for the Higgs mass (top-left), the neutralino relic density
(top-right) and the stau yield (bottom-left) in the $m_0$-$M_{1/2}$ plane, 
after requiring $\delta m < 0.1\GEV$ and a neutralino LSP\@.
In the bottom-right plot we show values for the lightest stau mass
($m_{\stau_1}$) after the additional constraints on $m_h$ and
$Y^0_{\stau}$ have been imposed (see text for details). }
\label{fig:planeConst}
\end{figure}

For the results presented below we scan (with 14\,k points) over the
ranges:
\begin{eqnarray*}
-42000\GEV < & A_0 & < -1000\GEV,\\
630\GEV < & M_{1/2} & < 1100\GEV \mbox{ and }\\
144\GEV < & m_0 & < 463\GEV
\end{eqnarray*}
with $\tan\beta = 10$ and $\mu > 0$.
The requirements on the stau-neutralino mass splitting drastically
restricts the allowed CMSSM parameter space.
In Fig.~\ref{fig:planeConst} we show the $m_0$-$M_{1/2}$ plane along with the
corresponding values for the Higgs mass and $\omg$. All the
points shown satisfy $\delta m < 0.1\GEV$ and contain a neutralino LSP\@.
As we can see, at the right
edge of the points shown, the Higgs mass falls below 120\,GeV, while for
$M_{1/2} \gtrsim 1$\,TeV, the neutralino relic density violates Planck's upper
bound.
Furthermore, to the left of the points shown (low $m_0$), the stau becomes the
LSP\@. We also show values for the stau relic abundance, computed according to
Eqs.~\ref{eq:ytau} and \ref{eq:yneu}. As we can see all points allow for the
minimum stau yield required to provide a solution to the $^7$Li-problem.
Finally, we show in the bottom-right plot values for the lightest stau mass
after imposing all the minimal conditions listed above. As shown, the
constraints on the Higgs mass and the relic density (as well as $\delta m$)
imply $200\GEV < m_{\stau_1} < 460\GEV$.

\subsubsection*{LHC Constraints}

After identifying the region of the CMSSM parameter space consistent with {\it
Planck} ($\omg < 0.1283$), the Higgs mass and the solution to the $^7$Li-problem
we proceed to discuss the constraints from MET and HSCP searches at the LHC.
As described in Sec.~\ref{sec:sms}, we have modified the public version of
\smodels to include the CMS search for HSCPs.
Within this framework we can simultaneously apply the MET and HSCP constraints
to the CMSSM parameter space.
While the MET constraints directly make use of the upper limits on the
production cross sections (for a given simplified model) provided by ATLAS and
CMS, the HSCP constraints use the efficiency maps for the CMS exotic
search~\cite{Khachatryan:2015lla}, as described in Sec.~\ref{sec:effGen}.
Since the cascade decays of the SUSY particles in the scenario considered here
may end up either on the lightest stau or on the lightest neutralino, we expect both
the MET and the HSCP searches to be relevant for the parameter space shown in
Fig.~\ref{fig:planeConst}.
Therefore we simultaneously apply both constraints and we consider
a point in the parameter space excluded if, {\it for at least one} of  the MET
or the HSCP searches, the signal cross section for a given topology (or sum of
topologies in the case of HSCP searches), $\sigma_\text{th}$, is higher than its 
corresponding experimental upper limit ($\sigma_\text{UL}$).

\begin{figure}[th]
\centering
\setlength{\unitlength}{1\textwidth}
\begin{picture}(1,0.375)
 \put(-0.02,-0.0){ 
  \put(0.04,0.005){\includegraphics[width=0.475\textwidth]{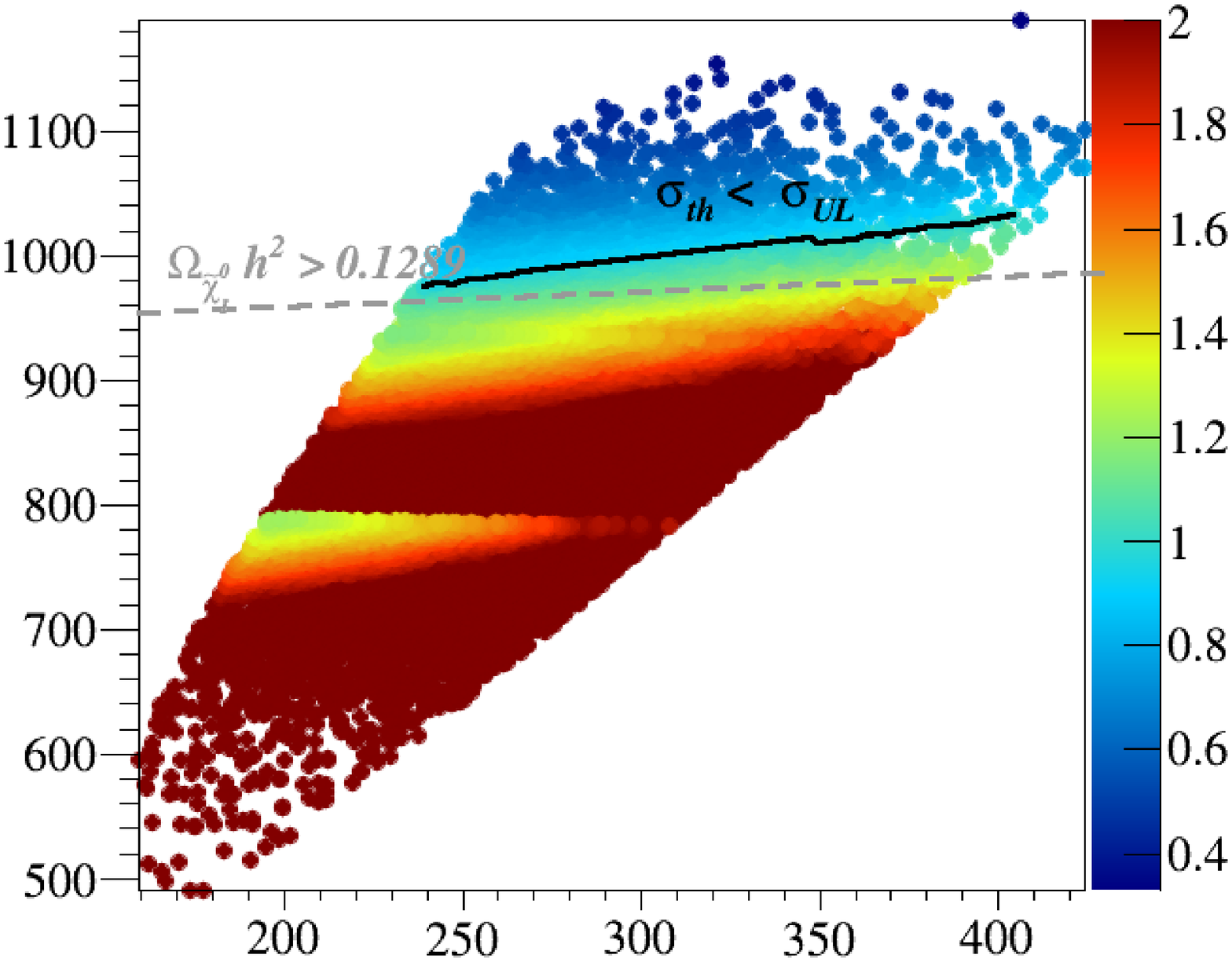}}
  \put(0.04,0.142){\rotatebox{90}{\footnotesize $M_{1/2}\,[\text{GeV}]$}}
  \put(0.23,0.0){\footnotesize $m_0 \,[\text{GeV}]$}
  \put(0.482,0.23){\rotatebox{-90}{\footnotesize $\sigma_\text{th}/\sigma_\text{UL}$}}
  }
 \put(0.48,-0.0){ 
  \put(0.04,0.005){\includegraphics[width=0.475\textwidth]{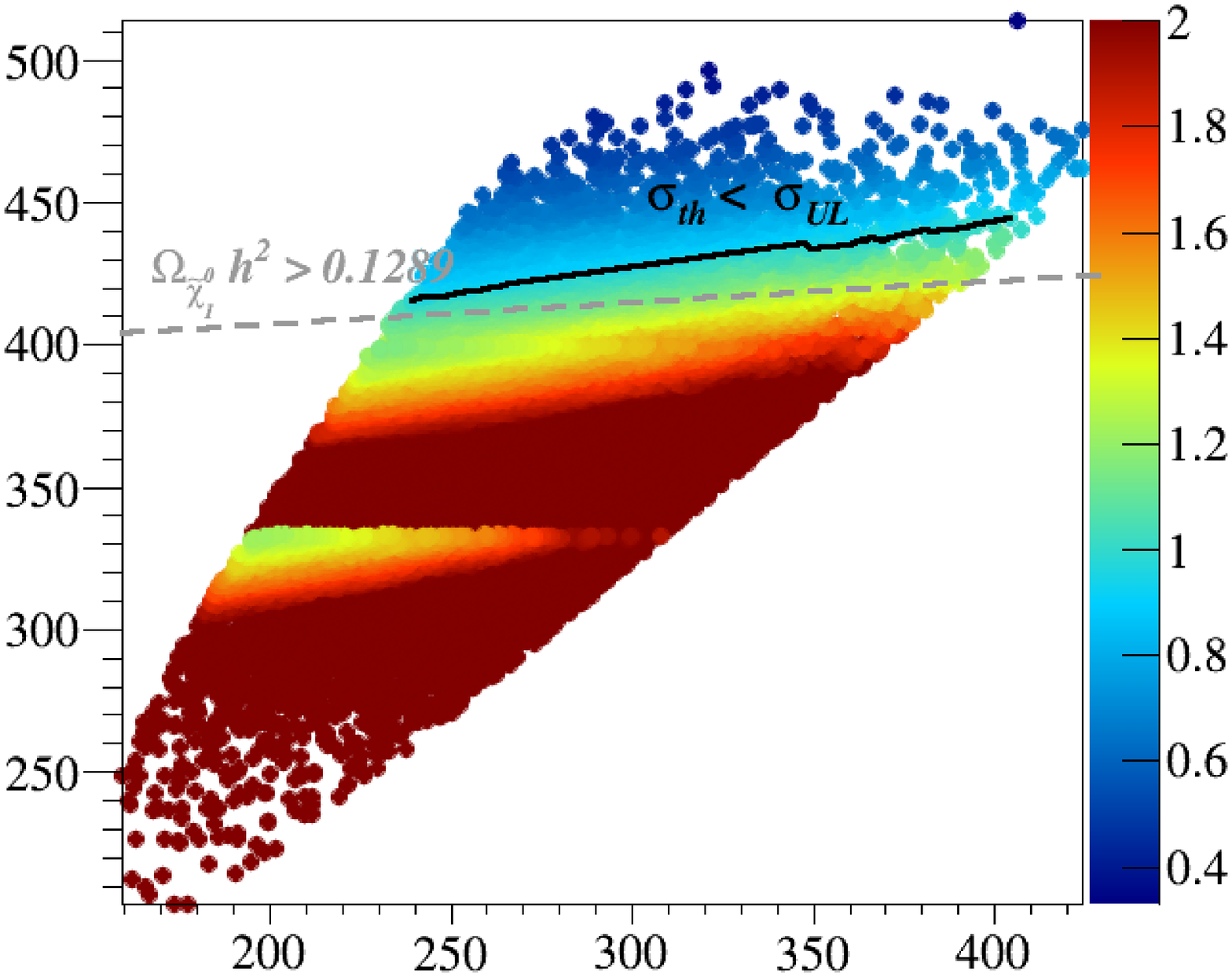}}
  \put(0.04,0.147){\rotatebox{90}{\footnotesize $\mstau\,[\text{GeV}]$}}
  \put(0.23,0.0){\footnotesize $m_0 \,[\text{GeV}]$}
  \put(0.482,0.23){\rotatebox{-90}{\footnotesize $\sigma_\text{th}/\sigma_\text{UL}$}}
}
\end{picture}
\caption{Values for the signal cross section ($\sigma_\text{th}$) over the experimental 
95\% CL upper limit ($\sigma_\text{UL}$) in the $m_0$-$M_{1/2}$ (left) and
$m_0$-$\mstau$ (right) planes. Points with $\sigma_\text{th}/\sigma_\text{UL} > 1$ 
are excluded by either MET searches or HSCP searches at the LHC\@.}
\label{fig:excl}
\end{figure}

In Fig.~\ref{fig:excl} we show the $\sigma_\text{th}/\sigma_\text{UL}$
ratio\footnote{Since here we are simultaneously considering HSCP and
MET constraints, we only show the $\sigma_\text{th}/\sigma_\text{UL}$ ratio for the most
constraining analysis (maximum ratio).} in the $m_0$-$M_{1/2}$ and $m_0$-$\mstau$
planes, as well as the lines for the upper limit on $\omg$ (dashed gray) and  
$\sigma_\text{th}/\sigma_\text{UL} = 1$ (solid black).
Values up to $M_{1/2} \simeq 1\TEV$ can be excluded, which 
corresponds to $\mstau \simeq 450\GEV$.
We also notice that the excluded region (where $\sigma_\text{th}/\sigma_\text{UL} > 1$)
extends to higher $M_{1/2}$ values when $m_0$ reaches its highest values (right
edge). In this region the stop mass is suppressed due to large $|A_0|$ values
and 
the pair production of stops is considerably enhanced, thus resulting in higher
signal cross sections.
The CMS paper~\cite{Khachatryan:2015lla} provides a conservative
limit on the stau mass, $\mstau > 260\GEV$, which is based on exclusive stau
pair production and is fairly model independent. Here, however, we can derive a
much more stringent bound, since we are able to consider the inclusive
production of staus from production of heavier sparticles and their
decays.\footnote{As already mentioned, since we only have computed the
efficiencies for the finite number of simplified models listed in
Tabs.~\ref{tab:defModels1} and~\ref{tab:defModels2}, all topologies not included
in the $\mathcal{M}1$-$\mathcal{M}8$ models or falling outside our mass grid do
not contribute to $\sigma_\text{th}$.} 

We can also compare the
bound derived here with the one obtained by CMS for the inclusive stau
production in the GMSB scenario (see Ref.~\cite{Khachatryan:2015lla} for
details). In the GMSB case the inclusive production (for a given $\mstau$)
is smaller than in the CMSSM scenario discussed above, due to the
presence of heavier squarks and gluinos. Nonetheless, CMS quotes $\mstau > 500\GEV$,
which is higher than the one found here for the CMSSM scenario.
This is mainly due to the fact that, while 100\% of the GMSB signal considered by
CMS goes to HSCP final states, a considerable fraction of our (CMSSM) signal goes into final
states with missing energy (neutralino final states), thus reducing the reach of
HSCP searches. Furthermore, the signal efficiencies for the events containing
one or two HSCPs are smaller for the CMSSM than for the GMSB scenario,
as a result of the different spectra as well as the fact that most of the events 
in the CMSSM signal contain only one HSCP\@. Nonetheless, we are still able 
to exclude all the region of parameter space (for $\tan\beta=10$) consistent with
{\it Planck's} upper bound on the Dark Matter relic abundance. Therefore we
conclude that, for low values of $\tan\beta$, the solution to the $^7$Li within
the CMSSM is no longer compatible with the LHC and Dark Matter constraints.

Finally, we comment on the feature appearing around $M_{1/2} \simeq
750\GEV$ or $\mstau \simeq 320\GEV$ in Fig.~\ref{fig:excl}. As discussed in
Sec.~\ref{sec:effGen}, for the HSCP searches we consider four signal regions 
($\text{SR}_0$, $\text{SR}_{100}$, $\text{SR}_{200}$ and $\text{SR}_{300}$). 
For $\mstau < 334\GEV$ the
efficiencies for $\text{SR}_{200}$ and $\text{SR}_{300}$ are taken as zero,
so the parameter space is constrained by the upper limits for
$\text{SR}_{100}$. Once $\mstau > 334\GEV$ ($M_{1/2} > 800\GEV$),
the efficiencies for $\text{SR}_{200}$ are no longer zero and this signal
region becomes the most constraining one, as shown by the sharp transition 
seen in Fig.~\ref{fig:excl}. This transition, however, does not affect
our results, since all the points in this region are excluded.

\begin{figure}[t]
\centering
\setlength{\unitlength}{1\textwidth}
\begin{picture}(1,0.435)
 \put(-0.02,-0.0){ 
  \put(0.04,0.002){\includegraphics[width=0.48\textwidth]{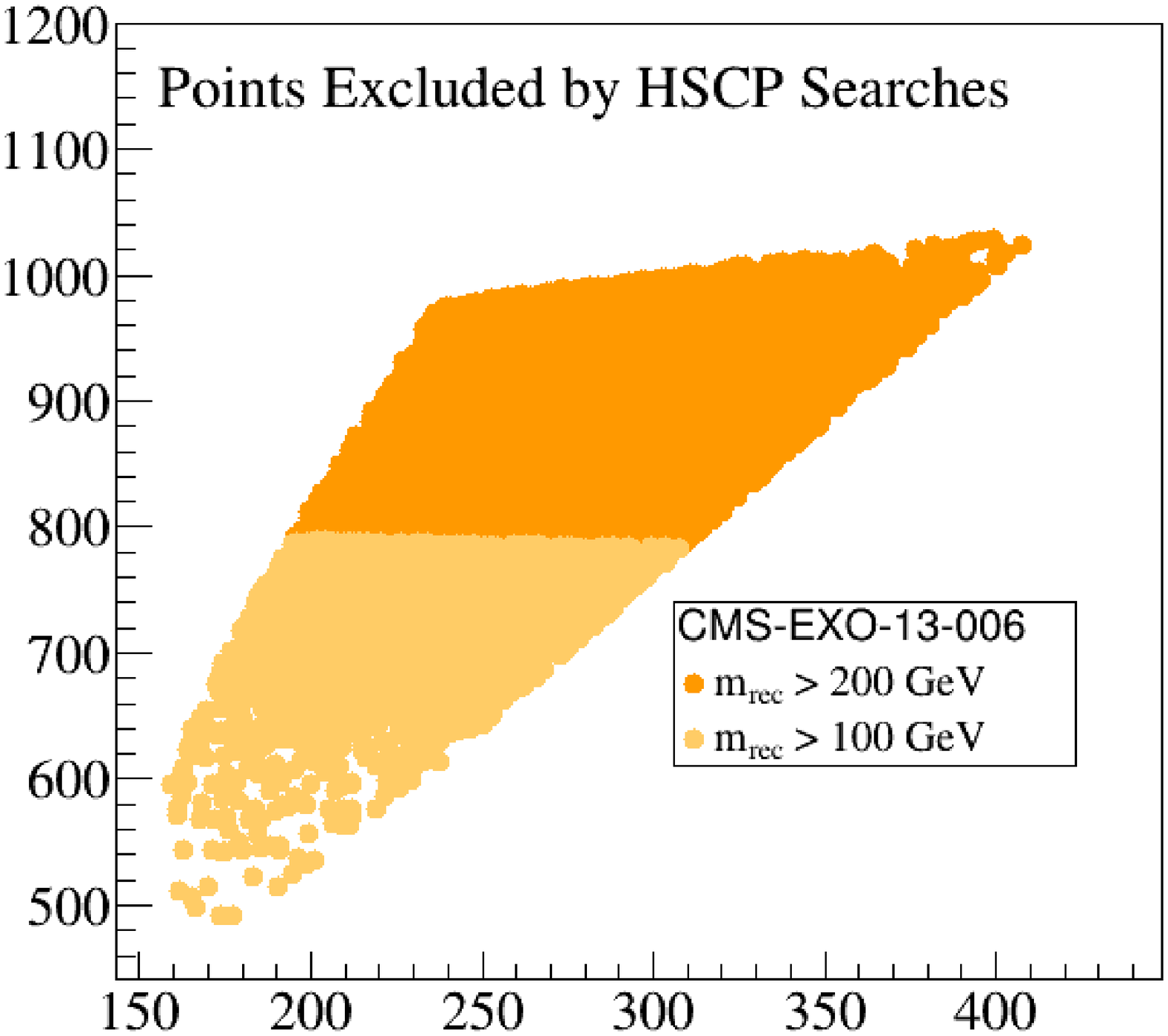}}
  \put(0.04,0.165){\rotatebox{90}{\footnotesize $M_{1/2}\,[\text{GeV}]$}}
  \put(0.245,0.0){\footnotesize $m_0 \,[\text{GeV}]$}
  }
 \put(0.48,-0.0){ 
  \put(0.04,0.002){\includegraphics[width=0.48\textwidth]{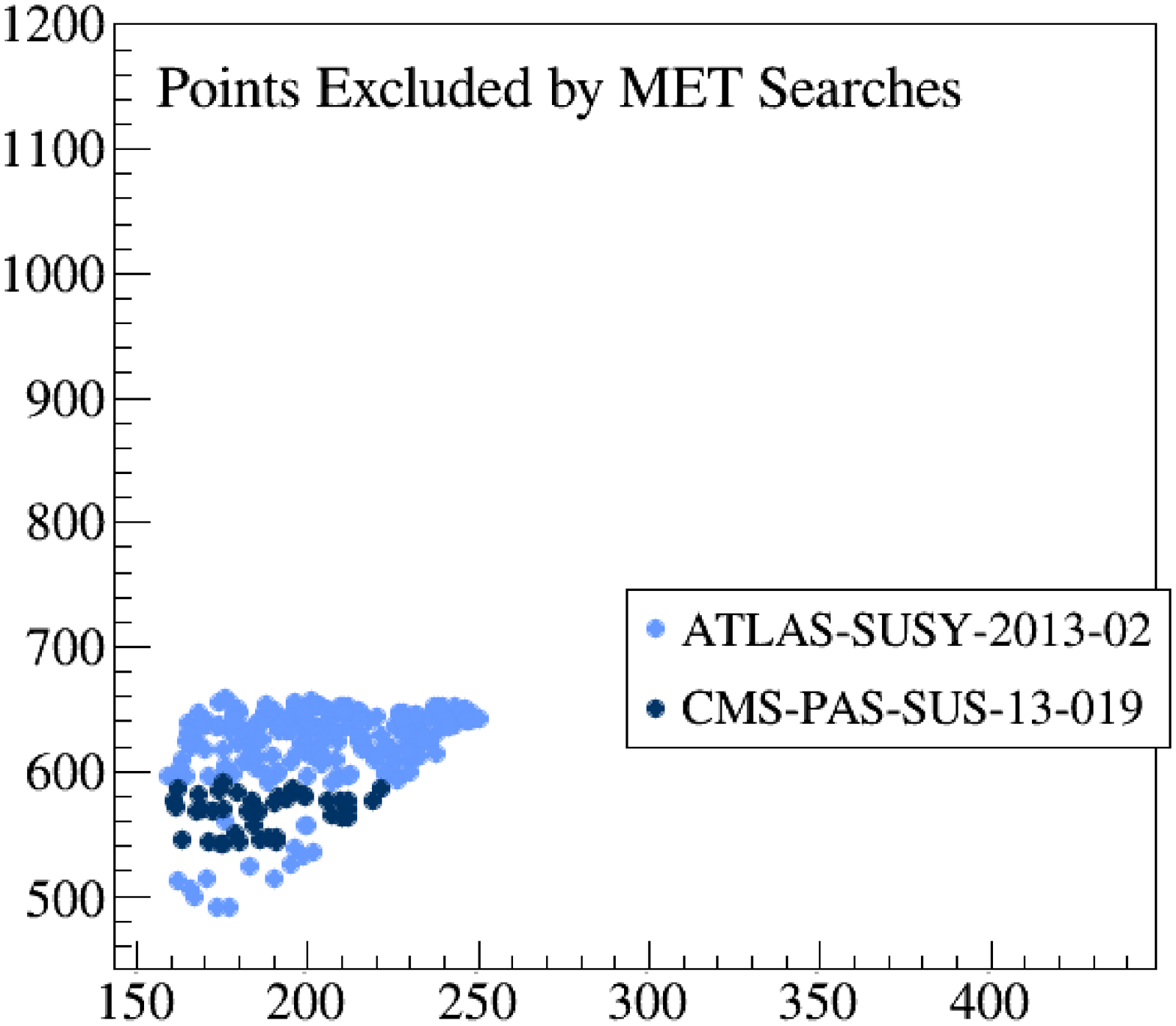}}
  \put(0.04,0.165){\rotatebox{90}{\footnotesize $M_{1/2}\,[\text{GeV}]$}}
  \put(0.245,0.0){\footnotesize $m_0 \,[\text{GeV}]$}
}
\end{picture}
\caption{Points excluded at 95\% CL by HSCP (left) and MET (right) searches 
in the $m_0$-$M_{1/2}$ plane. The distinct signal regions for the HSCP search from
CMS-EXO-13-006~\cite{Khachatryan:2015lla} are shown as light orange ($\text{SR}_{100}$)
and dark orange ($\text{SR}_{200}$). Signal regions $\text{SR}_{0}$ and $\text{SR}_{300}$
were also considered but are less constraining than $\text{SR}_{100}$ and $\text{SR}_{200}$ 
for this model.
For the MET searches we show by distinct colors
the constraints from CMS~\cite{Khachatryan:2015vra} (dark blue) and 
ATLAS~\cite{Aad:2014wea} (light blue) analyses.}
\label{fig:anas}
\end{figure}

Since in the model considered here the signal cross section splits into a HSCP
signal and a MET signal, we expect both the MET and the HSCP searches to have a
smaller reach than in a scenario where the signature is pure MET or pure HSCP\@.
In order to compare the reach of MET searches against the one of HSCP searches,
we show in Fig.~\ref{fig:anas} the most constraining (the one with the
highest $\sigma_\text{th}/\sigma_\text{UL}$ ratio) HSCP analysis (left) and MET
analysis (right).
As we can see, the constraints from MET searches exclude points up to $M_{1/2}
\simeq 650\GEV$, which corresponds to $m_{\tilde{g}} \simeq 1500\GEV$ or
$m_{\tilde{q}} \simeq 1350 \GEV$. The most constraining topologies in this case
are simply squark pair production followed by direct decay to the neutralino LSP\@.
On the other hand, the HSCP searches can exclude up to $M_{1/2}
= 1\TEV$ or $m_{\tilde{g}} = 2250 \GEV$ and $m_{\tilde{q}} = 2050\GEV$. As
expected, the HSCP constraints allows us to exclude a much larger region of the 
parameter space.

\subsection{Comparison with Full Simulation}

As already mentioned, the method outlined in Sec.~\ref{sec:sms} used to
obtain the results above relies on a few approximations. First, several details
of the full model are neglected, such as the spin of the intermediate particles, 
the production channel ($t$-channel or $s$-channel), kinematical effects due 
to off-shell decays and others. Second, when computing the signal cross sections
($\sigma_\text{th}$) for the full model we can only include the simplified models
$\mathcal{M}1$-$\mathcal{M}8$ for which we have computed efficiencies.
Finally other small effects such as the interpolation of the efficiency map
grid and the effect of neglecting topologies with weights below
$\tilde{\sigma}_{\min}$
can also affect the final result obtained through the simplified models
approach. Therefore it is relevant to compare the results obtained in
Sec.~\ref{sec:results} with a full Monte Carlo simulation.
In order to make this comparison we select $\mathcal{O}(100)$ representative points from
the CMSSM scan discussed in Sec.~\ref{sec:results} and compute their
signal cross sections for each of the HSCP signal regions
performing the full simulation via \textsc{Pythia}~6 followed by the analysis 
detailed in Sec.~\ref{sec:effGen}.

\begin{figure}[b]
\centering
\setlength{\unitlength}{1\textwidth}
\begin{picture}(0.65,0.45)
 \put(-0.02,-0.03){ 
  \put(0.04,0.01){\includegraphics[width=0.64\textwidth]{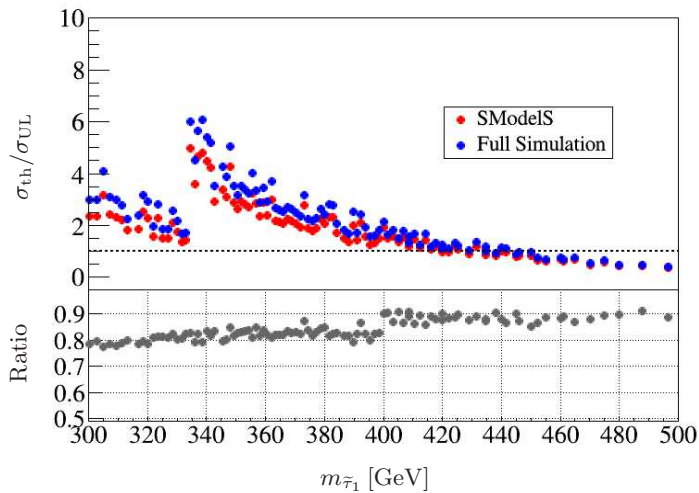}}
  \put(0.035,0.28){\rotatebox{90}{\footnotesize $\sigma_\text{th}/\sigma_\text{UL}$}}
  \put(0.035,0.12){\rotatebox{90}{\footnotesize Ratio}}
  \put(0.305,0.03){\footnotesize $\mstau \,[\text{GeV}]$}
  }
\end{picture}
\caption{Comparison between the results obtained with the full simulation (blue) and the modified 
\textsc{SModelS} version (red) for the HSCP search. In the upper frame we show the ratios between 
the signal cross section ($\sigma_\text{th}$) and the 95\% CL upper limit ($\sigma_\text{UL}$) for the 
two methods, whilst the lower frame shows their ratio, i.e. 
$(\sigma_\text{th}/\sigma_\text{UL})_\text{Full} / (\sigma_\text{th}/\sigma_\text{UL})_\text{SModelS}$.}
\label{fig:comp}
\end{figure}

In Fig.~\ref{fig:comp} we show the ratio $\sigma_\text{th}/\sigma_\text{UL}$ for the
best signal region as a function of the stau mass obtained from the modified
\textsc{SModelS} version (red points) and from the full simulation (blue points).
The lower frame shows the ratio between $\sigma_\text{th}/\sigma_\text{UL}$ for
the full simulation and \textsc{SModelS}. As shown in the figure,
the agreement is within $\sim$ 20\%
for all of the mass range shown.
As already mentioned, since we do not compute efficiencies for all possible
simplified models, we expect the excluded region obtained by \textsc{SModelS}
to be conservative. This is indeed what is seen in Fig.~\ref{fig:comp}, where
the \textsc{SModelS} value for $\sigma_\text{th}$ is always below the one
obtained with the full simulation. In order to guide the eye we also show
$\sigma_\text{th}/\sigma_\text{UL} = 1$ as a dashed line, hence all points above
this line are excluded. Even though \textsc{SModelS} is conservative, both
\textsc{SModelS} and the full simulation exclude stau masses up to $\simeq
450\GEV$, as already found in Sec.~\ref{sec:results}.

\section{Conclusions} \label{sec:concl}

Heavy stable charged particles (HSCPs) provide a prominent signature at the LHC
and are present in several well motivated BSM scenarios.
Most of the current experimental searches for HSCPs present their results
for specific BSM models, thus making it difficult to apply them to other
scenarios of interest.
Here we have re-interpreted these results in terms of simplified models,
what allowed us to derive constraints to a wide range of arbitrary BSM models
not considered previously.
To this end we have presented a new method to systematically decompose full BSM
models as a coherent sum of simplified models containing both stable and
quasi-stable particles. To this end we have defined a set of eight
simplified models containing one or two HSCPs in the final states and computed
the corresponding signal efficiencies as a function of the model parameters. 
With the inclusion of both the new decomposition method and the
efficiency maps to the program package \textsc{SModelS} we are able to apply
both MET and HSCP constraints to arbitrary BSM models containing a $\mathbb{Z}_2$ 
symmetry.

We showed that HSCP constraints on full BSM models can be reliably
applied through the simplified model framework presented here.
The constraints obtained by \textsc{SModelS} on the signal cross-sections for
the scenario studied here agreed within $\sim20\%$ 
with the full Monte Carlo simulation. These differences are similar to
other theo\-re\-ti\-cal uncertainties and, when translated to constraints
on the sparticles masses, do not lead to any significant difference between
the full simulation and \textsc{SModelS} results. Therefore we conclude
that the simplified models introduced here along with the
\textsc{SModelS} tools are well suited for confronting full models with
experimental HSCP searches.

We then applied our modified \textsc{SModelS}
program to the CMSSM stau co-annihilation strip, particularly considering the 
case of a nearly mass degenerate stau and neutralino.
In this part of the parameter space the stau becomes long-lived providing the HSCP
signature and presenting a 
potential solution to the Lithium problem. As the decay chains following the 
production of heavier SUSY particles can terminate in either the neutralino or
the stau, we encounter in this scenario both MET and HSCP signatures.
We have shown that the MET constraints for single simplified model topologies
allow us to exclude points up to $\mstau = 275\GEV$ (or $m_{\tilde{g}} \simeq
1500\GEV$ and $m_{\tilde{q}} \simeq 1350 \GEV$). On the other hand, using the
efficiency maps computed here, the HSCP searches can exclude up to $\mstau =
450\GEV$ (or $m_{\tilde{g}} = 2250 \GEV$ and $m_{\tilde{q}} = 2050\GEV$).
For small $\tan\beta$ values, the HSCP searches exclude the
whole parameter space consistent with a potential solution to the Lithium 
problem and {\it Planck's} bound on the neutralino relic abundance.

 \subsection*{Acknowledgements}
 
We are grateful to the SModelS group for useful discussions. This work
was supported by the German Federal Ministry of Education and Research 
(BMBF), by Funda\c{c}\~ao de Amparo \`a Pesquisa do Estado de S\~ao Paulo
(FAPESP) and by the Fonds de la Recherche Scientifique (FNRS).

\addcontentsline{toc}{section}{References}
\bibliographystyle{utphys.bst}
\bibliography{HSCP_ref}

\end{document}